\documentclass[journal]{IEEEtran}
\usepackage{mathrsfs}
\usepackage{pifont}
\usepackage{bbding}
\usepackage{tipa}
\usepackage{color}
\usepackage{cite}
\usepackage{epsfig}
\usepackage{graphicx}
\usepackage{url}
\usepackage{amsfonts}
\usepackage{multicol}
\usepackage{float}
\usepackage{times}
\usepackage{psfrag}
\usepackage{subfigure}
\usepackage{stfloats}
\usepackage{amsmath}
\usepackage{array}
\usepackage{stmaryrd}
\usepackage{amssymb}
\usepackage{epic}
\usepackage{graphicx}
\usepackage{curves}
\usepackage{balance}
\usepackage{algorithm}
\usepackage{algorithmic}
\usepackage{multirow}
\usepackage{amsmath}
\usepackage{xcolor}
\usepackage{graphics}
\usepackage{epstopdf}
\usepackage{enumerate}
\usepackage{geometry}
\usepackage{bm}
\usepackage{bbm}
\usepackage{dsfont}

\usepackage{caption}

\floatname{algorithm}{Algorithm}

\newtheorem{lemma}{Lemma}
\newtheorem{definition}{Definition}

\newtheorem{theorem}{Theorem}

\begin{document}

\title{Dynamic Trajectory and Power Control in Ultra-Dense UAV Networks: A Mean-Field Reinforcement Learning Approach }

\author{Fei Song, Zhe Wang, Jun Li, Long Shi, Wen Chen, Shi Jin

\thanks{Fei Song and Long Shi are with the School of Electronic and Optical Engineering, Nanjing University of Science Technology, Nanjing 210094, China (e-mail: fei.song@njust.edu.cn, slong1007@gmail.com).
Zhe Wang is with the School of Computer Science and Engineering, Nanjing University of Science and Technology, Nanjing 210094, China (email: zwang@njust.edu.cn).
Jun Li and Shi Jin are with the National Mobile Communications Research Laboratory, Southeast University, Nanjing 210096, China (e-mail: jleesr80@gmail.com, jinshi@seu.edu.cn).
Wen Chen is with the Department of Electronic Engineering, Shanghai Jiao Tong University, Shanghai 200240, China (e-mail: wenchen@sjtu.edu.cn).}

}

\date{}

\maketitle

\vspace{-0.5cm}
\begin{abstract}
  In ultra-dense unmanned aerial vehicle (UAV) networks, it is challenging to coordinate the resource allocation and interference management among large-scale UAVs, for providing flexible and efficient service coverage to the ground users (GUs).
  In this paper, we propose a learning-based resource allocation scheme in an ultra-dense UAV communication network, where the GUs' service demands are time-varying with unknown distributions.
  We formulate the non-cooperative game among multiple co-channel UAVs as a stochastic game, where each UAV jointly optimizes its trajectory, user association, and downlink power control to maximize the expectation of its locally cumulative energy efficiency under the interference and energy constraints. To cope with the scalability issue in a large-scale network, we further formulate the problem as a mean-field game (MFG), which simplifies the interactions among the UAVs into a two-player game between a representative UAV and a mean-field. We prove the existence and uniqueness of the equilibrium for the MFG, and propose a model-free mean-field reinforcement learning algorithm named maximum entropy mean-field deep Q network (ME-MFDQN) to solve the mean-field equilibrium in both fully and partially observable scenarios. The simulation results reveal that the proposed algorithm improves the energy efficiency compared with the benchmark algorithms. Moreover, the performance can be further enhanced if the GUs' service demands exhibit higher temporal correlation or if the UAVs have wider observation capabilities over their nearby GUs.
\end{abstract}


  \section{Introduction}
  With the commercialization of fifth generation (5G) and beyond wireless communication technology, unmanned aerial vehicles (UAVs) mounted base stations~\cite{EE_UAV,UAV_NOMA,A_survey_ch} can be deployed to enhance the coverage~\cite{Optimal_3D,EE_3D,Wireless_C} and reliability~\cite{UAV_commun_5G} of the existing terrestrial infrastructure in hot spots, outdoor events, and remote areas. Given the advantages of mobility, flexibility, and line-of-sight (LoS) propagation, UAVs have great potential to support wireless communications \cite{TA_UAV}, crowdsensing~\cite{HP_UAV,DS_EE,UAV_MTA}, mobile edge computing (MEC)~\cite{SC_UAV}, and caching~\cite{Multi_UAV_Traj} services for ground users~(GUs).

  For the UAV-aided communication networks, it is challenging to jointly optimize the UAV trajectory and real-time resource allocation in a dynamic and unknown environment.
  Deep reinforcement learning (DRL) algorithms provide solutions to sequential decision-making problems, e.g., UAV path planning, power control, and bandwidth allocation, in a trial-and-error manner without prior knowledge of the environmental distributions.
  In a multi-UAV network, it is non-trivial to coordinate the decisions among the intelligent UAV agents in a decentralized way~\cite{UAV_traj_opt,D_EE,MUAV_TP}. Multi-agent reinforcement learning~(MARL) algorithms have been recently applied in multi-UAV networks for the joint design of the UAV trajectory and resource allocation, aiming to minimize the energy consumption~\cite{EE_UAV}, maximize the energy efficiency~\cite{EE_3D}, or minimize the task execution delay~\cite{Collaborative_comp}.
  Compared with the single-UAV networks~\cite{Path_plan,3D_UAV,3D_Deploy}, there are basically two major challenges for resource allocation in the multi-UAV networks.
  On the one hand, each UAV has a limited coverage range and sensing capability, resulting in incomplete observation of the global network states. On the other hand, the environment faced by each UAV is non-stationary since the evolution of its local performance depends on not only the dynamic environment but also the aggregate interference from other intelligent UAVs.

  The resource allocation problem for multi-UAV networks can be briefly divided into two categories: cooperative and non-cooperative scenarios.
  For the fully cooperative multi-UAV network, the joint resource optimization among the agents can be formulated as a decentralized partially observable Markov decision process (Dec-POMDP), where all agents (e.g., UAVs) jointly maximize the social welfare of the entire network.
  To combat the partial observability and non-stationary problems, the centralized training and decentralized execution (CTDE) framework has been adopted in \cite{MADRL_traj,MADRL_task,MARL_res_man}, where each agent uploads its local information to a centralized controller for global training purposes, and then executes the local actions according to the learned policy in a decentralized manner.
  For the non-cooperative multi-UAV network, the agents (e.g., UAVs) may belong to different operators and be interested in optimizing their local performance only.
  In this scenario, the interactions among multiple self-interested UAVs can be formulated as a stochastic game, where each UAV makes its local resource decisions and path planning to maximize its long-term data rate~\cite{MARL_res_all}, energy efficiency~\cite{MA_coll}, or minimize the communication delay~\cite{Interference_Man}. Nash equilibrium characterizes an equilibrium point of the joint policy profile, where each agent acts according to its best response to the strategic behavior of others.
  Considering the sequential nature of stochastic game, it is usually costly to solve the Nash equilibrium, especially in a dynamic and unknown environment. Due to the privacy concerns, it may be difficult for the self-interested agents to find a trusted training central, and thus the MARL algorithms based on the CTDE framework in~\cite{MADRL_traj,MADRL_task,MARL_res_man} may not be applicable to non-cooperative scenarios. In \cite{MA_coll} and \cite{Interference_Man}, the authors adopt the fully decentralized MARL algorithms to solve the Nash equilibrium across the non-cooperative UAV agents without the aid of the centralized controller.
  Usually, the fully decentralized MARL algorithm suffers from the threat of non-stationarity due to the uncoordinated interactions and interference among agents.
  To deal with this issue, the UAV agents in \cite{MA_coll} and \cite{Interference_Man} exchange information (e.g., local state) with each other for the joint training of their policies, which however, may also cause privacy issues and excessive communication overhead.

  Mean-field game (MFG) effectively formulates the stochastic game among the massive number of agents in the large-scale networks.
  Based on the mean-field theory, MFG simplifies the interactions among multiple agents in the stochastic game into a two-player game between a representative agent and a mean-field term that characterizes the aggregated behavior of massive opponents~\cite{Downlink_power}.
  Mean field games have been applied in multi-UAV edge networks to reduce the user download delay~\cite{Delay_opt,Multi_UAV_delay}, minimize the energy consumption~\cite{MFGT}, or maximize the energy efficiency~\cite{Massive_UAV,Joint_p_control}. In~\cite{Delay_opt,Multi_UAV_delay,MFGT,Massive_UAV,Joint_p_control}, the mean-field equilibrium among multi-UAV agents is obtained by iteratively solving the Hamilton-Jacobi-Bellman (HJB) and Fokker-Planck-Kolmogorov (FPK) equations.
  Specifically, HJB equation models the representative agent's interaction with the mean-field, and FPK equation describes the evolution of the mean-field term.
  However, the formulations of HJB and FPK functions rely on the exact statistical modeling of the environment, and it is infeasible to solve the mean-field equilibrium in a highly dynamic and unknown environment.
  The recent advances in~\cite{MF_MARL,Appr_sol,LMFG} utilize the mean-field reinforcement learning (MFRL) algorithms to solve the mean-field equilibrium via the model-free methods.
  The authors in~\cite{Downlink_Trans} solve the mean-field equilibrium for downlink power control in ultra-dense non-cooperative UAV networks based on the reinforcement learning (RL) algorithms.
  In~\cite{MFRL}, a mean-field trust region policy optimization (TRPO) algorithm is proposed to solve the mean-field equilibrium for large-scale UAV deployment, aiming at maximizing the fairness of multi-UAV communication coverage under energy cost constraints. However, the aforementioned literature assumes that the GUs' service demands are static across time, and has not yet addressed the time-varying GUs' demands. Intuitively, if the GUs' demands change over time, the UAV can either adjust its trajectory or increase its transmit power to enhance its local service throughput.
  On the one hand, adjusting the UAV's location closer to the active user can improve the communication quality, but at the higher cost of flying time and energy consumption.
  Our recent work in~\cite{COW} has addressed this issue by proposing a RL-based adaptive UAV deployment policy catering to the time-varying GUs' demands in a single-UAV network.
  On the other hand, in a multi-UAV network, increasing the transmit power is more time and energy efficient to enhance the UAV coverage, but with a threat to increase the mutual interference among the UAVs. Therefore, it is important to jointly design the transmit power and UAV trajectory to balance the tradeoff between the service cost and interference, which is especially challenging to analyze for large number of UAV agents.

  In this work, we study the non-cooperative resource allocation in a large-scale UAV communication network with time-varying GUs' demands. We model the interactions among the ultra-dense UAVs as an MFG, and propose an MFRL algorithm to solve the mean-field equilibrium. The main contributions of this work are given as follows.
  \begin{itemize}
    \item We propose a non-cooperative resource sharing scheme in a downlink UAV communication network, where multiple solar-powered UAVs serve as self-organizing aerial base stations to provide communication services to the GUs with dynamic service demands.
        We formulate the non-cooperative game among the co-channel UAVs as a stochastic game, where each UAV jointly optimizes its transmit power, user association and trajectory to maximize the expectation of its own cumulative energy efficiency under the interference and energy constraints.
    \item To tackle with the scalability issue in a large-scale network, we further extend the stochastic game into an MFG. The interactions among the large number of individual agents in a stochastic game are simplified into a two-player game between a representative agent and a mean-field term that characterizes the aggregated behavior of a large number of UAV opponents. We prove the existence and uniqueness of the mean-field equilibrium as a solution to the MFG, and propose a two-step method to obtain the mean-field equilibrium by iteratively updating the optimal policy of the representative UAV agent and the mean-field distribution of the other UAVs until convergence.
    \item We propose a model-free MFRL algorithm named maximum entropy mean-field deep Q network (ME-MFDQN) to approximate the mean-field equilibrium based on the above two-step approach. Specifically, we adopt maximum entropy to enable the agent to balance between the exploitation of the locally optimal policy and the exploration over the other policy candidates, which prevents the algorithm from trapping into the local optima. Moreover, we extend the algorithm to approximate the mean-field equilibrium in the partially observable scenario, where each UAV has limited observability over the nearby GUs' dynamic demands.
    \item The simulation results show that the proposed ME-MFDQN algorithm can effectively improve the average energy efficiency of the representative UAV compared with the benchmark algorithms.
        Moreover, we observe an improvement in energy efficiency when the UAVs are more certain about the environment, i.e., if the GUs' service demands exhibit a higher correlation across time, or if the UAVs have a wider observation radius.
        In these cases, it shows that the UAVs would explore less new service opportunities to reduce the travel costs and transmit at lower power to avoid excessive mutual interference.

  \end{itemize}

  The rest of the paper is organized as follows. In Section~II, the system model of large-scale non-cooperative UAV network is introduced. In Section~III, we formulate the non-cooperative interactions among UAV agents as a stochastic game and extend it to an MFG, then we derive a two-step method to solve the mean-field equilibrium. In Section~IV, the model-free MFRL algorithms are proposed to approximate the mean-field equilibrium for fully and partially observable scenarios. Section~V discusses the simulation results. Finally, Section~VII concludes the paper.

  \section{System Model}

  We consider an ultra-dense UAV network comprising a set of $ \mathcal{K} =\{1, 2, \ldots, K\}$ rotary-wing UAVs to support GUs in $K$ cells as shown in Fig.~1.
  Each UAV $k \in \mathcal{K}$ provides the downlink communication services to a set of~$\mathcal{U}_k = \{1,\ldots, U\}$ GUs by adaptively flying across its served cell~$k$ at a certain altitude $d$ according to GUs' dynamic service demands.
  Without loss of generality, each cell $k$ is divided into $U$ sectors, where GU $u\in \mathcal{U}_k$ is located at the center of the sector.
  For each UAV $k$, it adapts its location to the time-varying GUs' demands by visiting one of its hovering points $n\in \mathcal{U}_k$ inside serving cell~$k$, where we assume each hovering point $n$ is right above GU~$u$.

\begin{figure}[t]
 \centering
    \includegraphics[width=3.6in,angle=0]{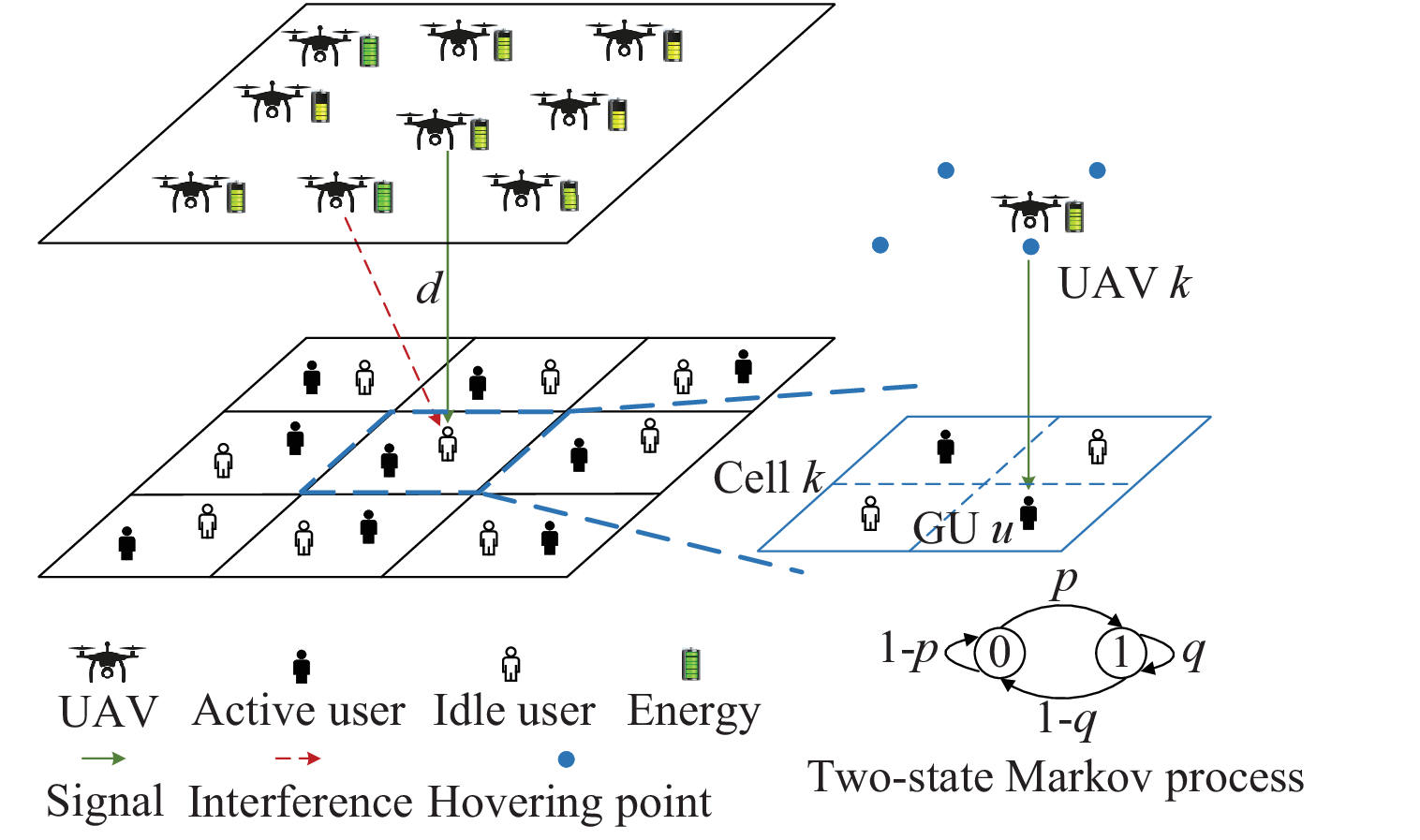}
    \caption{An ultra-dense UAV network provides communication services to GUs with time-varying service demands.}
\end{figure}

  Consider a time-slotted system, where the duration of each time slot $t$ is $\tau$.
  We denote the service demand state of GU $u$ at time slot $t$ by $x_{u}(t)\in\{0,1\}$.
  We assume the time-varying service demand of each GU $u$ follows a two-state Markov process \cite{COW}: state $x_{u}(t)=1$ represents the active mode for requesting the communication services, and state~$x_{u}(t)=0$ represents the idle mode.
  As shown in Fig.~1, the state transition probabilities from state $x_{u}(t)$ to next state $x_{u}(t+1)$ are given by $\Pr (x_{u}(t+1) = 1|x_{u}(t) = 0) = p$, $\Pr (x_{u}(t+1) = 1|x_{u}(t) = 1) = q$, $\Pr (x_{u}(t+1) = 0|x_{u}(t) = 0) = 1-p$ and $\Pr (x_{u}(t+1) = 0|x_{u}(t) = 1) = 1-q$, respectively.
  We assume that UAV $k$ can observe the service demand states of all GUs within its serving cell $k$.
  In Section~V-B, we will extend the discussions to the partially observable scenario that each UAV can only access to the real-time service demand state of some nearby GUs and does not know the true state of the rest of GUs in its cell.

  \subsection{Fly-Hover-Communicate Protocol }

  We divide the duration of each time slot into two parts of~$\tau_1$ and $\tau_2$, as shown in Fig.~2.
  At the beginning of each time slot, the UAV jointly decides its service location and user association.
  Denote the UAV's selected hovering location at time slot $t$ by $n_k(t) \in \mathcal{U}_k$. If $n_k(t)=n_k(t-1)$, UAV~$k$ will keep hovering at the current location, otherwise the UAV will fly to another hovering point during $\tau_1$.
  We adopt the fly-hover-communication protocol~\cite{Energy_Min}, which has two modes.
  For mode $\text{1}$, the UAV flies to a new hovering point during $\tau_1$ and then provides communication services to its associated GU during $\tau_2$.
  For mode $\text{2}$, the UAV stays at the current hovering point for data transmission during the entire $\tau$.
  We consider that the UAV can associate with any GU inside its service cell which may not necessarily be the one below its hovering point.
  Without loss of generality, we assume the UAV serves only one GU during each time slot, and it does not transmit during its flight unless staying at a certain hovering point.

  \begin{figure}[H]
 \centering
    \includegraphics[width=2.4in,angle=0]{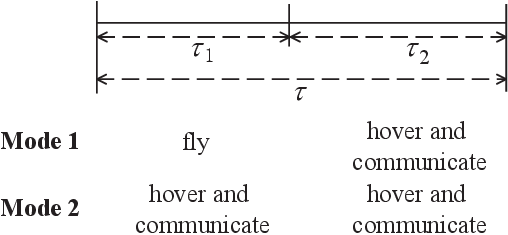}
    \caption{Division of a time slot.}
\end{figure}

  \subsection{Communication Model }
  The channel coefficient $h_{k,u}(t)$ between UAV $k$ and GU~$u$ for time slot $t$ can be expressed as
  \begin{align}
  h_{k,u}(t) = \sqrt{{l}_{k,u}(t)}g_{k,u}(t),
  \end{align}
  where $g_{k,u}(t)$ is the small-scale fading, ${l}_{k,u}(t)$ is the path loss characterizing the large-scale fading. The large-scale fading \cite{Energy_Min} is modeled as a random variable depending on the LoS probability $Pr^{\rm{LoS}}_{k,u}(t)$ and non-LoS (NLoS) probability~$1-Pr^{\rm{LoS}}_{k,u}(t)$.
  At time slot $t$, $Pr^{\rm{LoS}}_{k,u}(t)$ between UAV $k$ and GU $u$ is given by
  \begin{align}
  Pr^{\rm{LoS}}_{k,u}(t) = \frac{1}{1+c_1\exp(-c_2(\theta_{k,u}(t)-c_1))},
  \end{align}
  where $c_1$ and $c_2$ are the parameters depending on the propagation environment, and $\theta_{k,u}(t)$ is the elevation angle between the two nodes. We therefore express ${l}_{k,u}(t)$ as
  \begin{align}
  {l}_{k,u}(t)=\left\{
  {\begin{array}{*{20}{l}}
  \!\!A^{\rm{L}} (D_{k,u}(t))^{-\alpha^{\rm{L}}},\!\!\!\!
  &\text{with}\;Pr^{\rm{LoS}}_{k,u}(t),      \\
  \!\!A^{\rm{NL}} (D_{k,u}(t))^{-\alpha^{\rm{NL}}},\!\!\!\!
  &\text{with}\;1\!\!-\!\!Pr^{\rm{LoS}}_{k,u}(t),
  \end{array}}\right.
  \end{align}
  where $A^{\rm{L}}$ and $A^{\rm{NL}}$ are the path loss at the reference distance for LoS and NLoS links \cite{DEC_FRL}, respectively. Moreover,~$\alpha^{\rm{L}}$ and $\alpha^{\rm{NL}}$ respectively represent the path loss exponents for LoS and NLoS links, and $D_{k,u}(t)$ is the distance between UAV $k$ and GU $u$.

  In addition, we consider the small-scale fading $g_{k,u}(t)$ between UAV $k$ and GU $u$ for LoS and NLoS links following Nakagami-$m$ fading \cite{Down_cov}. The probability density function for Nakagami-$m$ fading is given by
  \begin{align}
  \label{nakagami}
  \mathcal{F}(g_{k,u}(t))\!=\!\! \left\{\!\!\!\!
  {\begin{array}{*{20}{l}}
  \frac{2m^m (g_{k,u}(t))^{2m\!-\!1}}{\Gamma(m)\Omega^m} \exp(\!-\frac{m(g_{k,u}(t))^2}{\Omega}),\\
  \;\;\;\;\;\;\;\;\;\;\;\;\;\;\;\;\;\;\;\;\;\;\;\;\;\;\;\;\;\;\;\;\;\;\;\;\;\;\text{LoS link},      \\
  \frac{2g_{k,u}(t)}{\Omega}\! \exp(\!-\frac{(g_{k,u}(t))^2}{\Omega}),\;\text{NLoS link},
  \end{array}}\right.
  \end{align}
  where $\Omega$ is the spread controlling factor, 
  $m$ is the Nakagami shape factor for LoS link, and $\Gamma(m)$ is the Gamma function.
  In addition, we consider $m=1$ for NLoS link that follows Rayleigh fading.

  For UAV $k$, the transmit power at time slot $t$ is $p_k(t) \in \{p^1, p^2, \ldots, p^{\rm{max}}\}$ and $p^{\rm{max}}$ is the maximum allowable transmit power of UAV.
  We assume full frequency reuse that all UAVs share the same frequency.
  At time slot $t$, the signal-to-interference-plus-noise-ratio (SINR) at GU $u$ during $\tau_1$ and $\tau_2$ are respectively expressed as
  \begin{align}
  {{\rm{SINR}}_{k,u}^1(t)} =  {\frac{\varphi_k(t) \delta_{k,u}(t) x_u(t) p_k(t) |h_{k,u}(t)|^2} {\sum_{k'\in \mathcal{K}\setminus k} I^1_{k'}(t) + N_0 }},
  \end{align}
  and
  \begin{align}
  {{\rm{SINR}}_{k,u}^2(t)} =  {\frac{ \delta_{k,u}(t) x_u(t) p_k(t) |h_{k,u}(t)|^2} {\sum_{k'\in \mathcal{K} \setminus k}  I^2_{k'}(t) + N_0 }},
  \end{align}
  where $\varphi_{k}(t)$ is the UAV deployment indicator at time slot~$t$ that returns $1$ if UAV $k$ keeps hovering at the current location (mode 2), and it returns $0$ if the UAV decides to fly to another hovering point (mode 1).
  Note that if~$n_k(t)=n_k(t-1)$, the UAV will stay hovering at the current location as in mode 2, i.e., $\varphi_{k}(t) = 1$.
  Likewise,~$\delta_{k,u}(t) \in \{0, 1\} $ denotes whether UAV $k$ is associated with GU $u$ at time slot~$t$, which returns $1$ if GU~$u$ is selected by UAV~$k$ for the communication service, otherwise it returns $0$. Thus, the user association decision for UAV $k$ is defined as $\bm{\delta}_k(t)=\{\delta_{k,u}(t)\}_{u \in \mathcal{U}_k}$.
  At each time slot, we assume each UAV can connect to at most one GU, i.e., $ \sum_{u\in \mathcal{U}_k} \delta_{k,u}(t) \leq 1$.
  Moreover, $N_0$ is the additive white Gaussian noise power, $I_{k'}^1(t)$ and $I_{k'}^2(t)$ denote the interference that GU $u$ receives from UAV $k'$ during $\tau_1$ and~$\tau_2$, respectively. Note that the interference depends on UAV $k'$'s decisions on deployment mode $\varphi_{k'}(t)$, user association $\delta_{k',u'}(t)$, transmit power $p_{k'}(t)$, and its associated GU $u'$'s activity state $x_{u'}(t)$. The expressions of $I_{k'}^1(t)$ and $I_{k'}^2(t)$ are respectively given by
  \begin{align}
  {I}_{k'}^1(t) = \sum_{u'\in \mathcal{U}_{k'}} \varphi_{k'}(t) \delta_{k',u'}(t) x_{u'}(t) p_{k'}(t) |h_{k',u}(t)|^2,
  \end{align}
  and
  \begin{align}
  {I}_{k'}^2(t) =  \sum_{u'\in \mathcal{U}_{k'}} \delta_{k',u'}(t) x_{u'}(t) p_{k'}(t) |h_{k',u}(t)|^2.
  \end{align}

  Consider that each UAV provides communication service to GU (if active) with a fixed rate, where GU can decode the message from the associated UAV if the received SINR is above the target threshold $\eta$.
  Hence, for UAV $k$, the communication is successful if its received SINR satisfies ${{\rm{SINR}}_{k}^1(t)} = \sum_{u\in \mathcal{U}_{k}} {{\rm{SINR}}_{k,u}^1(t)} \geq \eta$ and ${{\rm{SINR}}_{k}^2(t)} = \sum_{u\in \mathcal{U}_{k}} {{\rm{SINR}}_{k,u}^2(t)} \geq \eta$ during $\tau_1$ and $\tau_2$, respectively. Given the system bandwidth $B$, the achievable rate ${R_k(t)}$ of UAV $k$ at time slot $t$ is expressed as (\ref{rate}) at the top of next page, where $\mathds{1}(\cdot)$ is an indicator function that returns~$1$ if the given condition is true, otherwise it returns~$0$.

  \begin{figure*}[htb]
  \begin{align}
  \label{rate}
   {R_k(t)}=\left\{
  {\begin{array}{*{20}{l}}
  \mathds{1}\left({\rm{SINR}}_{k}^2(t)\geq \eta\right)\tau_2 B{\log _2}\left( 1 + \eta \right) ,& \text{$\text{mode}$ $\text{1}$} ,      \\
  \left[\mathds{1}\left({\rm{SINR}}_{k}^1(t)\geq \eta\right)\tau_1 +\mathds{1}\left({\rm{SINR}}_{k}^2(t)\geq \eta\right)\tau_2\right]B{\log _2}\left( 1 + \eta \right), &\text{$\text{mode}$ $\text{2}$},
  \end{array}}\right.
  \end{align}
  \hrule
  \end{figure*}

  \subsection{Energy model}

  \subsubsection{Energy Harvesting Model}
  For sustainable energy supply, we consider that each UAV is equipped with a solar panel to harvest solar energy \cite{Optimal_3D}. During time slot $t$, the solar energy harvested by UAV $k$ is given by
  \begin{align}
  e^{\rm{s}}_k(t) = \varrho \vartheta G \exp\left({-\beta d^{\rm{cloud}}_k(t)}\right) \tau,
  \end{align}
  where $\varrho$ and $\vartheta$ are the energy harvesting efficiency and the equivalent area of the solar panels, respectively. The coefficient of $G$ represents the average solar radiation intensity, $\beta$ denotes the absorption coefficient of the cloud,and $d^{\rm{cloud}}_k(t)$ is the thickness of the cloud above UAV $k$ at time slot $t$.

  \subsubsection{Energy Consumption Model}
  The energy consumption of each UAV $k$ includes the propulsion energy and communication energy \cite{Energy_Min}.
  Denote the UAV communication circuitry power by $p^{\rm{c}}$. At time slot $t$, the communication energy of UAV $k$ can be modeled as
  \begin{align}
  e^{\rm{c}}_k(t)=\left\{
  {\begin{array}{*{20}{l}}
  \left({p}_k(t)+p^{\rm{c}}\right)\tau_2,&\text{$\text{mode}$ $\text{1}$} ,      \\
  \left({p}_k(t)+p^{\rm{c}}\right)(\tau_1+\tau_2),&\text{$\text{mode}$ $\text{2}$} .
  \end{array}}\right.
  \end{align}

  For the rotary-wing UAV, the propulsion power consumption includes the induced, parasite, and blade profile power. We denote the flying speed of UAV $k$ at time slot $t$ by $v_k(t)$, which is the ratio of the flying distance and $\tau_1$. Thus, given the flying speed $v_k(t)$, the propulsion power of UAV $k$ at time slot $t$ is expressed as
  \begin{align}
  &p_k^{\rm{f}}(t)=\! p^{\rm{f1}}\left(1+\frac{3(v_k(t))^2}{ \zeta^2 (\omega^{\rm{rad}})^2}\right) \nonumber \\
  &+\!p^{\rm{f2}}\left( \sqrt{1+\frac{\rho^2 (\omega^{\rm{area}})^2 (v_k(t))^4}{W^2}}- \frac{\rho \omega^{\rm{area}} (v_k(t))^2}{W} \right)^{1/2} \nonumber \\
  &+ \!\frac{1}{2}\varepsilon\rho \omega^{\rm{sol}} \omega^{\rm{area}} (v_k(t))^3,
  \end{align}
  where $p^{\rm{f1}} = \frac{\varpi}{8}\rho \omega^{\rm{sol}} \omega^{\rm{area}} \zeta^3 (\omega^{\rm{rad}})^3$ and $p^{\rm{f2}} = (1+\lambda)\frac{W^{3/2}}{\sqrt{2\rho \omega^{\rm{area}}}}$ respectively denote the blade profile power and induced power of the UAV in hovering state,
  $W$ is the weight of the UAV,
  $\zeta$ is the blade angular velocity,~$\varpi$ and~$\lambda$ are the profile drag coefficient and incremental correction factor to induced power.
  The coefficients of~$\omega^{\rm{rad}}$,~$\omega^{\rm{area}}$ and~$\omega^{\rm{sol}}$ are the rotor radius, rotor disc area and rotor solidity, respectively.
  In addition, the coefficients of $\varepsilon$ and~$\rho$ represent the fuselage drag ratio and the air density, respectively. If UAV $k$ decides to fly to another hovering point during $\tau_1$ with a speed $v_k(t)$ at time slot $t$, the flying energy consumption of UAV $k$ is given by
  \begin{align}
  e^{\rm{f}}_k(t) = p^{\rm{f}}_k(t) \tau_1.
  \end{align}

  Moreover, the power consumption for hovering at a certain location is a constant, which is given by
  \begin{align}
  p^{\rm{h}} = p^{\rm{f1}} +p^{\rm{f2}}.
  \end{align}
  Thus, the hovering energy consumption of UAV $k$ at time slot $t$ is given by
   \begin{align}
  e^{\rm{h}}_k(t) = \left\{
  {\begin{array}{*{20}{l}}
   p^{\rm{h}} \tau_2, &\text{$\text{mode}$ $\text{1}$} ,      \\
   p^{\rm{h}} \left(\tau_1+\tau_2\right), &\text{$\text{mode}$ $\text{2}$} .
  \end{array}}\right.
  \end{align}

  \subsubsection{Energy Queue Model}
  Each UAV carries a battery to store the harvested energy from the solar panels and provide power for communication and propulsion.
  We assume that~$e^{\rm{max}}$ is the maximum battery capacity, and all UAVs are initialized with a fully charged state.
  At time slot $t$, the battery energy level $e_k(t)$ should satisfy
  \begin{align}
  0 < e_k(t) \leq e^{\rm{max}}.
  \end{align}
  Based on the above discussions, the total energy consumption of UAV $k$ at time slot $t$ is
  \begin{align}
  &e^{\rm{total}}_k(t) \nonumber \\
  &\;\;\;=\left\{
  {\begin{array}{*{20}{l}}
  p^{\rm{f}}_k(t) \tau_1 +  \left(p^{\rm{h}}+p_k(t)+p^{\rm{c}}\right)\tau_2,&\text{$\text{mode}$ $\text{1}$} ,      \\
  \left(p^{\rm{h}}+p_k(t)+p^{\rm{c}}\right)\left(\tau_1+\tau_2\right),&\text{$\text{mode}$ $\text{2}$} .
  \end{array}}\right.
  \end{align}

  Moreover, the energy consumed by flying, hovering, and communication should be no more than the energy level~$e_k(t)$ at the beginning of time slot $t$, which is described as the energy causality constraint $e^{\rm{total}}_k(t)\leq e_k(t)$.
  The evolution of energy queue dynamics can be expressed as
  \begin{align}
  \label{queue}
  e_k(t\!+\!1) \!=\! \min\left\{\left[e_k(t)\!-\!e^{\rm{total}}_k(t)\right]^+ \!+\!e_k^{\rm{s}}(t), e^{\rm{max}}\right\},
  \end{align}
  where $x^+ \triangleq \max\{x,0\}$.

  We define the energy efficiency of UAV $k$ as the achievable rate per unit cost of energy consumption, i.e.,
  \begin{align}
  \label{TE}
  EE_k(t) =\frac{R_k(t)}{e^{\rm{total}}_k(t)}.
  \end{align}

  \vspace{-2mm}
  \section{Formulation of Mean-Field Game}
  In this section, we consider a non-cooperative and self-organizing scenario in which each UAV selfishly adapts its trajectory planning and power control to the GUs' demands, aiming to maximize the achievable throughput within its serving cell in an energy efficient way. There are two key features of this resource allocation problem.
  First, due to the temporal correlation of system dynamics (e.g., GUs' demands, UAV's battery states), the joint optimization of each UAV's location and transmission policy not only affects the current service quality in its serving cell, but also the unpredictable rewards in future time slots.
  Second, due to the mutual interference and limited information exchange among the UAV transmitters, each UAV's local service reward is not only influenced by its own decision but also by the uncertain amount of aggregate interference from other UAVs in the network with unknown locations and transmission policies. We formulate the sequential decision-making problem across the non-cooperative and mutually interfered UAVs by a stochastic game in Section~III-A, where each UAV strives to attain its best possible local reward without the awareness of the distribution of system dynamics. Then, we extend the formulation to the MFG in an ultra-dense UAV network in Section~III-B.

  \vspace{-3mm}
  \subsection{Stochastic Game}

  The discrete-time $K$-agent stochastic game for multi-UAV control is described as follows. The stochastic game is represented by a $6$-tuple~$(\mathcal{K}, \mathcal{S}_{\rm{tot}},\{\mathcal{A}_k\}_{k\in \mathcal{K}}, \{r_k\}_{k\in \mathcal{K}},$ $ P, \gamma)$, which are described as follows.
  \begin{itemize}
    \item Agent set $\mathcal{K}$: $\mathcal{K} = \{1,\ldots,K\}$ represents the set of densely-deployed UAV agents, where each UAV $k$ is a rational and strategic player that aims to maximize its own utility.
    \item State space $\mathcal{S}_{\rm{tot}}$: $\mathcal{S}_{\rm{tot}}=\bigcup_{k\in \mathcal{K}}\mathcal{S}_k$, where $\mathcal{S}_k$ represents all possible states for UAV $k$.
        At the time slot~$t$, the state $s_k(t) \in \mathcal{S}_k$ of UAV $k$ includes the service demand states $\bm{x}_k(t)= \{x_{1}(t),\ldots,x_{U}(t)\}$ of all GUs within cell $k$, its hovering point position~$n_k(t-1) \in \mathcal{U}_k$ at the end of the previous time slot, and the energy level~$e_k(t)$ at the beginning of this time slot, i.e.,~$s_k(t) = \{\bm{x}_k(t), n_k(t-1), e_k(t)\}$. The joint state of all UAVs is denoted by~$s_{\rm{tot}}(t) = \{s_k(t)\}_{k\in \mathcal{K}} \in  \mathcal{S}_{\rm{tot}}$.
    \item Action space $\{\mathcal{A}_k\}_{k\in \mathcal{K}}$: $\mathcal{A}_k$ represents all possible actions for UAV $k$.
        At time slot $t$, the action~$a_k(t) = \{\bm{\delta}_k(t), n_k(t), p_k(t)\} \in \mathcal{A}_k$ of agent~$k$ includes the user association decision $\bm{\delta}_k(t)$, the selected hovering location $n_k(t)$, and the transmit power $p_k(t)$.
    \item Reward function $\{r_k\}_{k\in \mathcal{K}}$: At time slot $t$, each agent~$k$ receives the immediate reward $r_k(t)$ expressed as
        \begin{align}
        \label{reward}
        r_k(t)&\!=\! EE_k(t)-\sigma p_k(t)(\varphi_k(t)\tau_1+ \tau_2) \nonumber\\
        &\;\;\;-\!\xi\max\left\{e_k^{\rm{total}}(t)+e_k^{\min}\!-\!e_k(t),0 \right\},
        \end{align}
        where $\sigma$ is the interference penalty factor imposed by the system for the purpose of interference control, and~$\xi$ is the energy penalty factor that penalizes the event if the remaining energy $e_k(t)-e_k^{\rm{total}}(t)$ at the end of the time slot $t$ (before the new energy arrival~$e_k^s(t)$) is below the energy alarm threshold $e_k^{\min}$.
    \item Transition probability $P$: $\mathcal{S}_{\rm{tot}} \times \mathcal{A}_1 \times \cdots \times \mathcal{A}_K \rightarrow \mathcal{P}(\mathcal{S}_{\rm{tot}})$ describes the stochastic evolution of states over time, where $\mathcal{P}(\mathcal{S}_{\rm{tot}})$ is the set of probability distributions for state space $\mathcal{S}_{\rm{tot}}$.
    \item Discount factor $\gamma$: The constant $\gamma \in [0,1]$ is the reward discount factor across time.
  \end{itemize}

  For agent $k$, the action selection policy at time slot $t$ is defined as $\pi_k (t) \!\!: \mathcal{S}_{\rm{tot}} \rightarrow \mathcal{P}(\mathcal{A}_k)$, where $\mathcal{P}(\mathcal{A}_k)$ is the set of probability distributions for action space $\mathcal{A}_k$. Let~$\bm{\pi}_{\rm{tot}}(t) = [\pi_1(t),\ldots,\pi_K(t)]$ denote the joint policy of all UAVs at time slot $t$. Thus, the joint policy sequence for discrete-time is denoted as $\bm{\pi}_{\rm{tot}} = \left\{\bm{\pi}_{\rm{tot}}(t)\right\}_{t=0}^{\infty}$. Given the initial state~$s_{\rm{tot}}$, the state-value function is defined as the expectation of the cumulative discounted reward of agent $k$ under the joint policy $\bm{\pi}_{\rm{tot}}$, i.e.,
  \begin{align}
  \label{vf}
  &V_k(s_{\rm{tot}}, {\bm{\pi}}_{\rm{tot}}) \nonumber \\
  &\;\;\;= \mathbb{E}\left[\sum \limits_{t=0}^\infty \gamma^t r_k(s_{\rm{tot}}(t), a_{\rm{tot}}(t)) \left|s_{\rm{tot}}(0)=s_{\rm{tot}} \right.\right],
  \end{align}
  where $a_{\rm{tot}}(t) \triangleq \left[a_1(t),\ldots,a_K(t)\right]$.

  In this stochastic game, each UAV aims to obtain an optimal policy $\bm{\pi}_{\rm{tot}}^*$ to maximize its value function.
  However, the optimization of $V_k(s_{\rm{tot}},{\bm{\pi}}_{\rm{tot}})$ not only depends on its own policy $\bm{\pi}_k = \left\{\pi_k(t)\right\}_{t=0}^{\infty}$ but also on the joint policy of all agents. We therefore define Nash equilibrium~$\bm{\pi}_{\rm{tot}}^*$ for the stochastic game as an equilibrium point that no agent has a profitable deviation from $\bm{\pi}_{\rm{tot}}^*$.
  \begin{definition}
  The Nash equilibrium for the stochastic game is defined as a particular joint policy ${\bm{\pi}}_{\rm{tot}}^* = [\bm{\pi}_1^*,\ldots,\bm{\pi}_K^*]$ that satisfies
  \begin{align}
  &V_k(s_{\rm{tot}}, {\bm{\pi}}_{\rm{tot}}^*) \!=\! V_k(s_{\rm{tot}}, {\bm{\pi}}_k^*, {\bm{\pi}}_{-k}^*) \!\geq \! V_k(s_{\rm{tot}}, {\bm{\pi}}_k, {\bm{\pi}}_{-k}^*),
  \end{align}
  for all $s_{\rm{tot}}\in \mathcal{S}_{\rm{tot}}$ and $k\in \mathcal{K}$,
  where ${\bm{\pi}}_{-k}^* = [{\bm{\pi}}_1^*,\ldots,{\bm{\pi}}_{k-1}^*,{\bm{\pi}}_{k+1}^*,\ldots,{\bm{\pi}}_K^*]$ is the joint policy of all agents except $k$.
  \end{definition}

  However, solving the Nash equilibrium for the stochastic game in the ultra-dense UAV network is difficult (if not impossible) due to the space-time dynamics and the curse of dimensionality. Each UAV needs to act strategically to the joint actions of all other agents, where the dimension of the joint actions grows proportionally as the number of agents increases. To deal with this scalability issue, we employ MFG to characterize the interactions among a large population of UAVs, which will be discussed in the following subsection.

  \vspace{-1mm}
  \subsection{Mean-Field Game}
  MFG is a special form of stochastic game with a large number of strategic agents.
  In MFG, the interactions among individual agents in a $K$-player stochastic game are simplified into a two-player game between a representative agent and a mean-field term that characterizes the aggregated behavior of a large number of opponents.
  For MFG, the number of agents should be adequately large, since the evolution of the mean-field becomes deterministic with~$K \rightarrow \infty$ limit.
  In addition, we assume all UAVs are homogeneous, indistinguishable and exchangeable, where each UAV has the same state space and action space with~$\mathcal{S}_k=\mathcal{S}$ and $\mathcal{A}_k=\mathcal{A}$, $\forall~k$.
  Then, we will focus on a representative UAV due to the homogeneity of UAVs.
  Specifically, we define the mean-field as follows.

  \begin{definition}
  For the representative UAV $k$, the mean-field at time slot $t$ is defined as the joint state-action probability distribution over all other UAVs, which is given by
  \begin{align}
  \mathcal{L}(t) = \mathop {\lim }\limits_{K \to \infty } \frac{1}{K}\sum\limits_{k'\in \mathcal{K}\setminus k}  \mathds{1}_{\{s_{k'}(t)=s, a_{k'}(t)=a\}}.
  \end{align}
  Besides that, the state distribution $\mu(t)$ is given by
  \begin{align}
  \mu(t) = \mathop {\lim }\limits_{K \to \infty } \frac{1}{K}\sum\limits_{k'\in \mathcal{K}\setminus k} \mathds{1}_{\{ s_{k'}(t)=s \}}.
  \end{align}

  \end{definition}

  With a slight abuse of notation, we denote the state~$s_k(t)$, action $a_k(t)$ and reward $r_k(t)$ of the representative agent by $s(t) \in \mathcal{S}$, $a(t) \in \mathcal{A}$, and $r(t)$ at time slot~$t$. The corresponding policy for the representative agent at time slot $t$ is defined as $\pi(t) \!\!:\mathcal{S}\times \mathcal{P}(\mathcal{S}\times\mathcal{A})\rightarrow \mathcal{P}(\mathcal{A})$.
  At time slot $t$, given the local state $s(t)$ and the state-action distribution $\mathcal{L}(t)$, the representative agent chooses the action $a(t)$ according to its policy $\pi(t)$, i.e., $a(t)\sim \pi(s(t),\mathcal{L}(t))$.
  Then, the reward~$r(s(t),a(t),\mathcal{L}(t))$ is obtained, and the state $s(t)$ will evolve according to the state transition~$P(s(t+1)|s(t),a(t),\mathcal{L}(t))$. The state-value function for the representative agent is given by
  \begin{align}
  V(s, \bm{\pi}, \bm{\mathcal{L}})\! =\! {\mathbb{E}}  \left[ \sum\limits_{t = 0}^\infty \gamma^t r(s(t),a(t),\mathcal{L}(t))|s(0)\! =\! s \right],
  \end{align}
  where $\bm{\pi}=\{\pi(t)\}_{t=0}^\infty$ and $\bm{\mathcal{L}}=\{\mathcal{L}(t)\}_{t=0}^\infty$ with $\mathcal{L}(t) \in  \mathcal{M}$, $\mathcal{M}:= \mathcal{P}(\mathcal{S}\times \mathcal{A})$ is the set of joint probability distributions of state and action spaces. The objective of the representative agent is to solve its optimal policy $\bm{\pi}^*$ as the best response to the other agents' policies in order to maximize its state-value function $V(s, {\bm{\pi}}, \bm{\mathcal{L}})$.
  We rename the Nash equilibrium among the ultra-dense UAVs in the MFG as the mean-field equilibrium, which is defined as follows.

  \begin{definition}
   The mean-field equilibrium is defined as the policy-distribution pair $({\bm{\pi}}^*, {\bm{\mathcal{L}}}^*)$ that satisfies
  \begin{enumerate}
    \item (Representative agent side) Given the fixed distribution~$\bm{\mathcal{L}}^*$, for any policy $\bm{\pi}$ and state $s \in \mathcal{S}$, it satisfies
    \begin{align}
    V(s, {\bm{\pi}}^*, \bm{\mathcal{L}}^*)\geq V(s, \bm{\pi}, \bm{\mathcal{L}}^*).
    \end{align}

    \item (Mean-field side) Under the fixed policy ${\bm{\pi}}^*$, the joint state-action distribution for each agent matches $ \mathcal{L}^*(t)$ for all $t\geq 0$, where the action selection and state update for each agent follow $a(t)\sim \pi^*(s(t), \mathcal{L}^*(t))$ and $s(t+1)\sim P(\cdot|s(t), a(t), \mathcal{L}^*(t))$, respectively.
  \end{enumerate}
  \end{definition}

  We further explain Definition~3 as follows.
  For the representative agent side, it obtains its best response policy~$\bm{\pi}^*$ given the fixed mean-field $\bm{\mathcal{L}}^*$.
  For the mean-field side, the joint state-action distribution for each agent matches $\bm{\mathcal{L}}^*$ given the same policy $\bm{\pi}^*$ is adopted at all agents.
  The mean-field equilibrium can be iteratively obtained based on the two-step diagram as illustrated in Fig.~3, which is further explained as follows.

  \begin{figure}[t]
    \centering
    \includegraphics[width=2.8in,angle=0]{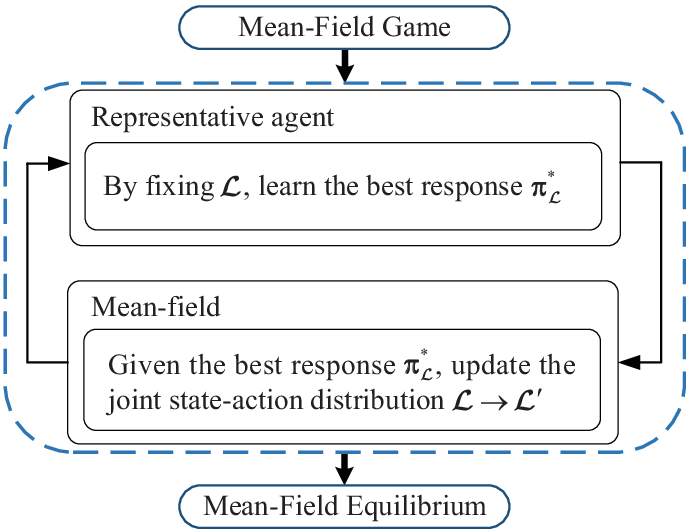}
    \caption{ The two-step iterative solution for MFG.}
  \end{figure}

  Step 1: Given the distribution $\bm{\mathcal{L}}$, the MFG reduces to a single-agent optimization problem, where the representative agent aims to find an optimal policy $\pi_{\bm{\mathcal{L}}}^*(t)\in \Pi :=\{\pi|\pi: \mathcal{S}\times \mathcal{P}(\mathcal{S} \times \mathcal{A}) \rightarrow \mathcal{P}(\mathcal{A})\}$ for maximizing its state-value function in (25).
  Define a mapping $\Upsilon_1$ from the fixed distribution $\bm{\mathcal{L}}$ to the solution $\bm{\pi}_{\bm{\mathcal{L}}}^*=\{\pi^*_{\bm{\mathcal{L}}}(t)\}_{t=0}^\infty$, which is given by
  \begin{align}
  \Upsilon_1: \left\{\mathcal{M}\right\}_{t=0}^{\infty} \rightarrow \left\{\Pi\right\}_{t=0}^{\infty}. 
  \end{align}
  We therefore can express the optimal policy as $\bm{\pi}_{\bm{\mathcal{L}}}^* = \Upsilon_1(\bm{\mathcal{L}})$. This sequence of $\bm{\pi}_{\bm{\mathcal{L}}}^*$ satisfies the representative agent side condition in Definition~3 given the mean-field side sequence ${\bm{\mathcal{L}}}$, where $V(s, \bm{\pi}_{\bm{\mathcal{L}}}^*, \bm{\mathcal{L}})\geq V(s, \bm{\pi}_{\bm{\mathcal{L}}}, \bm{\mathcal{L}})$ holds for any initial state $s$ and policy $\bm{\pi}_{\bm{\mathcal{L}}}$.

  Step 2: Given the optimal policy $\bm{\pi}_{\bm{\mathcal{L}}}^*$ obtained from Step~1, we update the distribution $\bm{\mathcal{L}}$ to a new distribution~$\bm{\mathcal{L}}'$.
  The state distribution induced by policy $\bm{\pi}^*_{\bm{\mathcal{L}}}$ is given by
  \begin{align}
  &\bm{\mu}' = \sum_{s\in \mathcal{S}} \bm{\mu} \sum_{a\in \mathcal{A}}\bm{\pi}^*_{\bm{\mathcal{L}}}P(s'|s,a, \bm{\mathcal{L}}).
  \end{align}
  Furthermore, the distribution is updated by $\bm{\mathcal{L}}' =  \bm{\mu}' \bm{\pi}^*_{\bm{\mathcal{L}}} $.
  Define a mapping $\Upsilon_2$ from the any distribution $\bm{\mathcal{L}}$ and any policy $\bm{\pi}$ to the another distribution $\bar{\bm{\mathcal{L}}}$, i.e.,
  \begin{align}
  \label{distribution}
  \Upsilon_2: \left\{\Pi \right\}_{t=0}^{\infty} \times \left\{\mathcal{M}\right\}_{t=0}^{\infty} \rightarrow \left\{\mathcal{M}\right\}_{t=0}^{\infty},
  \end{align}
  which is written as $ \bar{\bm{\mathcal{L}}} = \Upsilon_2(\bm{\pi}, \bm{\mathcal{L}})$.  Given distribution $\bm{\mathcal{L}}$ and the optimal policy $\bm{\pi}_{\bm{\mathcal{L}}}^*$, we can therefore express the new distribution as $\bm{\mathcal{L}}' = \Upsilon_2(\bm{\pi}_{\bm{\mathcal{L}}}^*,\bm{\mathcal{L}})$.

  Finally, we repeat the above two steps until $\bm{\mathcal{L}}'$ matches~$\bm{\mathcal{L}}$. Define a mapping  $\Upsilon: \left\{\mathcal{M}\right\}_{t=0}^{\infty} \rightarrow \left\{\mathcal{M}\right\}_{t=0}^{\infty} $ with~$\bm{\mathcal{L}}' = \Upsilon(\bm{\mathcal{L}}) = \Upsilon_2(\Upsilon_1(\bm{\mathcal{L}}), \bm{\mathcal{L}})$.
  Furthermore, we propose the following theorem to prove the existence and uniqueness of mean-field equilibrium.

  \begin{theorem}{(Existence and uniqueness of mean-field equilibrium).}
  With any initial $\bm{\mathcal{L}} \in \mathcal{M}$, the fixed point iteration $\bm{\mathcal{L}}' = \Upsilon_2(\Upsilon_1(\bm{\mathcal{L}}), \bm{\mathcal{L}})$ converges to the unique stationary mean-field equilibrium.
  \end{theorem}
  $Proof:$ See Appendix A . \hfill $\blacksquare$

  \section{Mean-Field Reinforcement Learning Algorithms}

  As discussed in the previous section, the two-step approach for solving the mean-field equilibrium depends on the exact statistical model of the transition probabilities, which may not be easy to obtain in practice. In this section, we propose the model-free MFRL algorithm to approximate the mean-field equilibrium in Section~IV-A. Subsequently, we extend the discussions to the partially observable scenarios in Section~IV-B.

  \subsection{Maximum Entropy Mean-field Reinforcement Learning Algorithm}
  Our proposed MFRL algorithm follows the main idea of the two-step approach. For the first step, with the fixed mean-field $\bm{\mathcal{L}}$, we find the optimal policy of the representative agent that maximizes the action-value function (Q function) $Q^{\bm{\pi}}$, i.e.,
  \begin{align}
  \label{pi}
  \bm{\pi}_{\bm{\mathcal{L}}}^* = \arg \max_{\bm{\pi}}  Q^{\bm{\pi}} (s, a, \bm{\mathcal{L}} ) ,
  \end{align}
  where the Q function $Q^{\bm{\pi}}$ is given by
  \begin{align}
  \label{Q_stan}
  Q^{\bm{\pi}} (s, a, \bm{\mathcal{L}} ) = {\mathbb{E}}  \left[ \sum\limits_{t = 0}^\infty \gamma^t r(t)|s(0) = s, a(0) = a \right],
  \end{align}
  with $r(t) \triangleq r(s(t),a(t), \bm{\mathcal{L}})$.

  It is important to note that obtaining the exact optimal Q value is infeasible due to the lack of statistical model, and the approximation via the conventional RL algorithms may result in the local optima if the objective function is multi-modal. We propose a novel maximum entropy MFRL algorithm named ME-MFDQN to address this issue. By incorporating the maximum entropy into the Q function, we redefine the Q function as~$Q_{\rm{ME}}^{\bm{\pi}}$, i.e.,
  \begin{align}
  \label{Q_me}
  &Q_{\rm{ME}}^{\bm{\pi}} (s, a, \bm{\mathcal{L}} ) \nonumber \\
  &\;=\! {\mathbb{E}} \left[\sum\limits_{t = 0}^\infty \gamma^l \left(r(t)\!+\!\phi \mathcal{H}(\pi(t))\right) \left|s(0) \!=\! s, a(0)\! =\! a \right.\right],
  \end{align}
  where $\phi$ is the weighting factor, and $\mathcal{H}(\pi(t))$ denotes the entropy of policy.
  In addition, $\pi(t) \triangleq \pi(a(t)|s(t), \bm{\mathcal{L}})$ is the policy indicating the probabilities of taking each action under state $s(t)$ and distribution $\bm{\mathcal{L}}$,
  and the entropy of policy~$\pi$ is given by $\mathcal{H}(\pi(t)) = -\log \pi(a(t)|s(t), \bm{\mathcal{L}})$, which measures the randomness of the policy. Different from \eqref{Q_stan}, we do not directly choose the policy that maximizes the expectation of the cumulative discounted reward, but also encourage explorations on other actions according to the entropy term. This prevents the agent from falling into the local optima due to repeatedly selecting the same action. Based on \eqref{Q_me}, we therefore express the representative agent's optimal policy $\bm{\pi}_{\bm{\mathcal{L}}}^*$ as
  \begin{align}
  \bm{\pi}_{\bm{\mathcal{L}}}^* = \arg \max_{\bm{\pi}}  Q_{\rm{ME}}^{\bm{\pi}} (s, a, \bm{\mathcal{L}} ).
  \end{align}

  Moreover, we denote the optimal Q function as~$Q^*_{\rm{ME}}(s,a,\bm{\mathcal{L}}) = \max_{\bm{\pi}} Q^{\bm{\pi}}_{\rm{ME}}(s,a,\bm{\mathcal{L}})$. Since it is difficult to directly derive the Q function, we utilize the neural network to approximate its value based on the DQN algorithm. We model the parameterized Q function $Q_{\rm{ME}}^{\psi}(s(t),a(t),\bm{\mathcal{L}})$ with parameter of $\psi$ and the parameterized target Q function~$\tilde{Q}_{\rm{ME}}^{\tilde{\psi}}(s(t),a(t),\bm{\mathcal{L}})$ with parameter of~$\tilde{\psi}$, respectively. The parameterized Q function~$Q_{\rm{ME}}^{\psi}(s(t),a(t),\bm{\mathcal{L}})$ can be expressed as
  \begin{align}
  &Q_{\rm{ME}}^{\psi}(s(t),a(t),\bm{\mathcal{L}}) \nonumber \\
  &\;\;= r(t)+ {\mathbb{E}}\left[\sum\limits_{l = 1}^\infty \gamma^l \left(r(t+l)\!+\!\phi \mathcal{H}(\pi(t+l)) \right)\right].
  \end{align}
  According to \cite{RL_DEP}, we further express the optimal policy of the representative agent as
  \begin{align}
  \label{pi_mfg}
  &\pi^*(a(t)|s(t), \bm{\mathcal{L}}) \nonumber \\
  &\;=\!\exp \left(\frac{1}{\phi}\left(Q_{\rm{ME}}^{\psi}(s(t),a(t), \bm{\mathcal{L}})\!-\! V_{\rm{ME}}^{\psi}(s(t),\bm{\mathcal{L}}) \right) \right),
  \end{align}
  where the parameterized state-value function is given by
  \begin{align}
  \label{v_theta}
  V_{\rm{ME}}^{\psi}(s(t), \bm{\mathcal{L}})\!=\! \phi \log \!\sum_{a \in \mathcal{A} } \!\exp\left(\frac{1}{\phi} {Q}_{\rm{ME}}^{\psi}(s(t), a, \bm{\mathcal{L}})\right).
  \end{align}

  During the training process, the gradient descent is used to update the network parameters by minimizing the loss function
  \begin{align}
  \label{loss}
  J_Q(\psi)\!=\! {\mathbb{E}}_{(s(t),a(t))\sim \mathcal{D}}\bigg[\frac{1}{2}&\Big( Q_{\rm{ME}}^{\psi}(s(t),a(t), \bm{\mathcal{L}})  \nonumber \\
  &-\!\tilde{Q}_{\rm{ME}}^{\tilde{\psi}}(s(t),a(t), \bm{\mathcal{L}}) \Big)^2 \bigg],
  \end{align}
  where $\mathcal{D}$ is a replay buffer (RB). The target Q function is given by
  \begin{align}
  &\tilde{Q}_{\rm{ME}}^{\tilde{\psi}}(s(t),a(t),\bm{\mathcal{L}}) \nonumber\\
  &\;\;= r(t)+ \gamma{\mathbb{E}}_{s(t+1)\sim \mathcal{D}}\left[\tilde{V}_{\rm{ME}}^{\tilde{\psi}}(s(t+1), \bm{\mathcal{L}})\right],
  \end{align}
  where the target state-value function $\tilde{V}_{\rm{ME}}^{\tilde{\psi}}(s(t+1), \bm{\mathcal{L}})$ is similar to \eqref{v_theta} by replacing $\psi$ with the target parameter $\tilde{\psi}$. The gradient of \eqref{loss} is given by
  \begin{align}
  \label{PSI}
  {\nabla}_{\psi}J_Q(\psi) =\Big(&Q_{\rm{ME}}^{\psi}(s(t),a(t), \bm{\mathcal{L}})-\gamma \tilde{V}_{\rm{ME}}^{\tilde{\psi}}(s(t+1), \bm{\mathcal{L}})\nonumber \\
  &- r(t)\Big)\nabla_{\psi}Q_{\rm{ME}}^{\psi}(s(t),a(t),\bm{\mathcal{L}}).
  \end{align}

  We summarize the proposed ME-MFDQN algorithm in Algorithm 1. In lines 1-3, we initialize the parameters and mean-field distribution. In lines 5-15, the optimal policy~$\bm{\pi}^*_{\bm{\mathcal{L}}}$ in Step 1 is obtained given the fixed mean-field distribution~$\bm{\mathcal{L}}$. Particularly, line 11 carries out the update of the Q function parameter $\psi$, and line 12 performs the update of the target parameter $\tilde{\psi}$. In line 16, we update the mean-field distribution $\bm{\mathcal{L}}$ in Step 2 according to $\bm{\pi}^*_{\bm{\mathcal{L}}}$ through the state transition. Finally, by repeating the above two steps until the new mean-field $\bm{\mathcal{L}}'$ matches $\bm{\mathcal{L}}$, we obtain the mean-field equilibrium of $({\bm{\pi}}^*, \bm{\mathcal{L}}^*)$.

  The training costs of the proposed ME-MFDQN algorithm mainly depends on the computational complexity.
  The total number of network layers in the Q network includes one input layer, $F$ hidden layers and one output layer. The number of neurons in the input layer includes the dimensions of the UAV's state and the mean-field term. First, the dimension of UAV's state includes the service demand states of all GUs within its cell, its hovering point position and the energy level, i.e., ${\rm{dim}}(s) = U_k+2$. Second, the dimension of mean-field term depends on the dimensions of state space ${\rm{dim}}(\mathcal{S}) = 2^{U_k}U_k L_{\rm{e}}$ and action space ${\rm{dim}}(\mathcal{A}) = U_k^2 L_{\rm{p}}$, where $L_{\rm{e}}$ and $L_{\rm{p}}$ are the number of energy levels and power levels for UAV, respectively. Thus, there are ${\rm{dim}}(s)+ {\rm{dim}}(\mathcal{S}){\rm{dim}}(\mathcal{A})$ neurons in the input layer. The number of neurons in $f$-th ($f \in \{1,\ldots,F\}$) hidden layer is $\Phi(f)$. The number of neurons in the output layer is the dimension of action space ${\rm{dim}}(\mathcal{A})$.
  Therefore, the computational complexity is expressed as ~$\mathcal{O}\big(\iota T \big[ \left({\rm{dim}}(s)+{\rm{dim}}(\mathcal{S}){\rm{dim}}(\mathcal{A})\right) \Phi(1)+$ $ \sum_2^{F}\Phi(f-1)\Phi(f) + \Phi(F){\rm{dim}}(\mathcal{A}) \big]\big)$, where~$\iota$ is the single weight's training computational complexity, and $T$ is the number of slots per episode.

  \begin{algorithm}
  \caption{Maximum Entropy MFDQN (ME-MFDQN)}
  \begin{algorithmic}[1]
    \STATE Initialize Q function parameter $\psi$
    \STATE Initialize parameter of target network $\tilde{\psi} = \psi$
    \STATE Initialize mean-field distribution $\bm{\mathcal{L}}_0$
    \FOR{$\kappa = 0, 1, \ldots$}
    \FOR{$\mathrm{episode} = 1, 2, \ldots$}
    \FOR{$t = 1, 2, \ldots$}
        \STATE Observe state $s$ and take action $a$ by \eqref{pi_mfg}
        \STATE Observe the reward $r$ and the next state $s'$
        \STATE Store experience $\langle s, a, r, s', \bm{\mathcal{L}}_\kappa \rangle$ in RB $\mathcal{D}$
        \STATE Sample random $I$ minibatch of experiences $\langle s, a, r, s', \bm{\mathcal{L}}_\kappa\rangle$ from $\mathcal{D}$
        \STATE Update $\psi$ according to (\ref{PSI})
        \STATE Update target parameter: $\tilde{\psi} \leftarrow \psi$
    \ENDFOR
    \ENDFOR
        \STATE Obtain the optimal policy $\bm{\pi}^*_{\bm{\mathcal{L}}_\kappa} $
        \STATE Compute the new distribution $\bm{\mathcal{L}}_{\kappa+1}$ based on $\bm{\pi}_{\bm{\mathcal{L}}_\kappa }^*$
    \ENDFOR
  \end{algorithmic}
  \end{algorithm}

  \subsection{Partially Observable Maximum Entropy Mean-field Reinforcement Learning Algorithm}
  In the previous subsection, we assume that the UAV can fully observe the GUs' service demands in its cell. However, in practice, the UAV may not be able to observe the real-time states of all GUs in its cell limited by its observability. According to this, we will consider a more practical scenario that the UAV can only observe the service demands from part of the GUs at the same time, which can be modeled as a partially observable stochastic game. Specifically, we define the partially observable stochastic game by a $7$-tuple~$(\mathcal{K}, \mathcal{S}, \{\mathcal{A}_k\}_{k\in \mathcal{K}},\{r_k\}_{k\in \mathcal{K}}, P, \gamma, \{\mathcal{O}_k\}_{k\in \mathcal{K}})$, where~$\mathcal{O}_k$ represents the observation space for UAV $k$.
  In the partially observable scenario, the local state $s_k(t)$ is not fully observed by agent $k$.
  At time slot $t$, the observation of UAV $k$ is defined as $o_k(t) = \{\bm{x}_k^{\rm{o}}(t), n_k(t-1), e_k(t)\} \in \mathcal{O}_k$, where $\bm{x}_k^{\rm{o}}(t) = \{x_{1}^{\rm{o}}(t),\ldots,x_{U}^{\rm{o}}(t)\}$ is the set of service demand observations of all GUs by UAV~$k$ within cell $k$ in the current time slot. Specifically, we set~$x_u^{\rm{o}}(t)\in \{0,1,2\}$ as UAV $k$'s observation on the service demand state of GU $u$, where $x_{u}^{\rm{o}}(t)=0$ represents GU $u$ is observed in active mode,~$x_{u}^{\rm{o}}(t)=1$ represents GU~$u$ is observed in idle mode, and $x_u^{\rm{o}}(t) = 2$ represents GU~$u$'s demand cannot be observed by UAV~$k$.
  Different from the fully observable scenario that only focuses on~$s_k(t)$, UAV $k$ in the partially observable scenario needs to record the entire sequence of the past trajectory $z_k(t)=\{o_k(t),o_k(t-1),\ldots,o_k(0)\}$ to infer the current state $s_k(t)$ and update the policy accordingly.
  At time slot $t$, the policy for agent $k$ is defined as~${\pi_k}(t) \!\!: \mathcal{Z}_k \rightarrow \mathcal{P}(\mathcal{A}_k)$, where $\mathcal{Z}_k$ is the history trajectory space and~$z_k(t) \in \mathcal{Z}_k$. However, it requires large storage capacity to record the entire past experiences which contains redundant information.

  In the ultra-dense UAV network, we extend the partially observable stochastic game to formulate a partially observable MFG (POMFG), which is regarded as the extension of the MFG discussed in Section~III-B into the partially observable scenario.
  With a slight abuse of notation, we denote the observation, history trajectory, action, and reward of the representative agent by~$o(t) \in \mathcal{O}$, $z(t) \in \mathcal{Z}$, $a(t) \in \mathcal{A}$ and $r(t)$.
  To reduce the input dimensionality of the entire observation history, we simplify the history trajectory at time slot~$t$ as~$z(t) =\{\hat{x}_{1}(t),\ldots,\hat{x}_U(t), \mathfrak{h}_1(t),\ldots,\mathfrak{h}_U(t), n(t-1), e(t)\}$, where~$\hat{x}_{u}(t)\in \{0,1\}$ represents the representative UAV's most recent observation on the idle $(0)$ or active~$(1)$ service demand state of GU $u$, $\mathfrak{h}_u(t)$ denotes the time elapsed since the most recent observation on this GU.
  The corresponding policy for the representative agent at time slot $t$ is redefined as ${\pi}(t)\!\!:\mathcal{Z}\times \mathcal{P}(\mathcal{O}\times\mathcal{A})\rightarrow \mathcal{P}(\mathcal{A})$.
  To solve this POMFG, we propose a partially observable ME-MFDQN that is similar to ME-MFDQN.
  Given the initial history trajectory $z$, the objective of POMFG is given by
  \begin{align}
  \bm{\pi}_{\bm{\mathcal{L}}^{\rm o}}^* \!=\!\arg\max \limits_{\bm{\pi}}{\mathbb{E}}\!\Bigg[\! \sum\limits_{t = 0}^\infty \gamma^t \!\left( r(t) \!\!+\! \phi \mathcal{H}\left(\pi(t)\right)\right)\!\left|z(0)\!\!=\! z \right. \!\!\Bigg],
  \end{align}
  where $r(t) \triangleq r(z(t),a(t),\bm{\mathcal{L}}^{\rm{o}})$, $\pi(t) \triangleq \pi(a(t)|z(t), \mathcal{L}^{\rm{o}})$ is the policy under the history trajectory $z(t)$ and the joint observation-action probability distribution $\bm{\mathcal{L}}^{\rm{o}}$.

  Thus, the agent is trained by minimizing the loss function
  \begin{align}
  J_Q(\psi)\!=\! {\mathbb{E}}_ {(z(t),a(t))\sim \mathcal{D}}\!\bigg[ \frac{1}{2} &\Big(Q_{\rm{ME}}^{\psi} (z(t),a(t), \bm{\mathcal{L}}^{\rm{o}})  \nonumber \\
  &-\!\tilde{Q}_{\rm{ME}}^{\tilde{\psi}}(z(t),a(t), \bm{\mathcal{L}}^{\rm{o}})\! \Big)^2\bigg].
  \end{align}
  The gradient is thus given by
  \begin{align}
  \label{PSI_o}
  {\nabla}_{\psi}J_Q(\psi) =\Big(&Q_{\rm{ME}}^{\psi}(z(t),a(t), \bm{\mathcal{L}}^{\rm{o}})-\gamma V_{\rm{ME}}^{\tilde{\psi}} \left(z(t+1), \bm{\mathcal{L}}^{\rm{o}}\right)\nonumber\\
  &- r(t)\Big) \nabla_{\psi}Q_{\rm{ME}}^{\psi}(z(t),a(t),\bm{\mathcal{L}}^{\rm{o}}).
  \end{align}

  The rest of the algorithm is similar to Algorithm 1, which is omitted here due to the lack of space.

  \section{Simulation Results}

  \begin{figure*}[t]
	\centering
	\subfigbottomskip=2pt
    \subfigure[]{
		\includegraphics[width=2.2in]{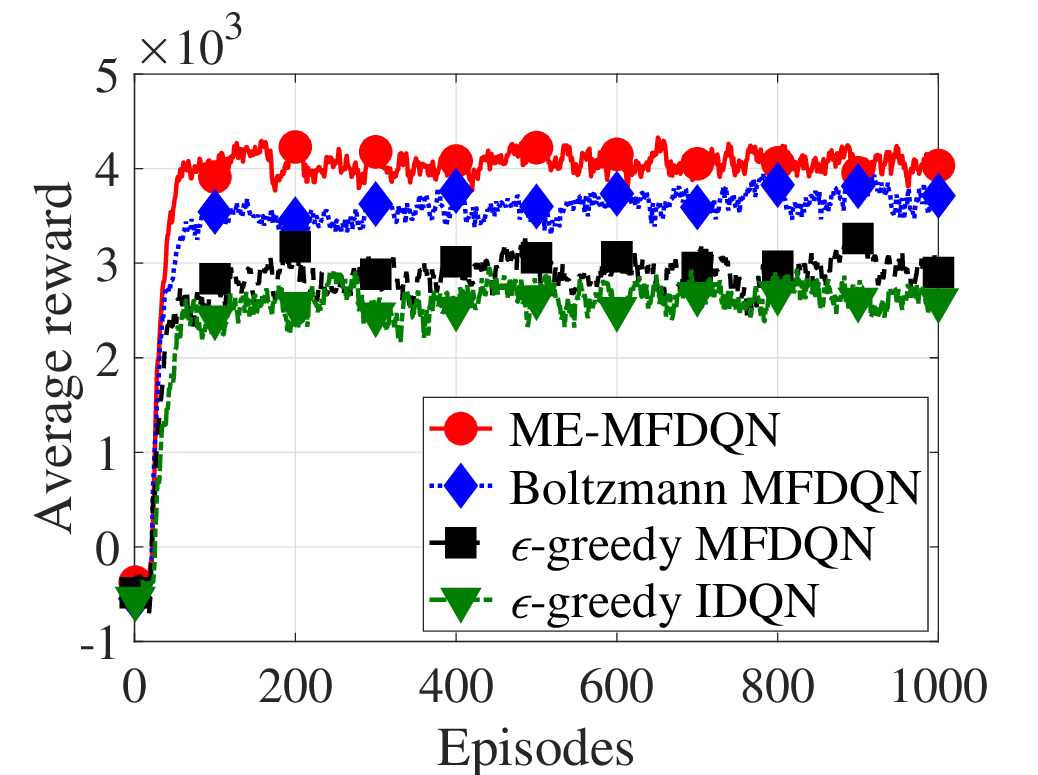}}\hspace{-3mm}
	\subfigure[]{
		\includegraphics[width=2.2in]{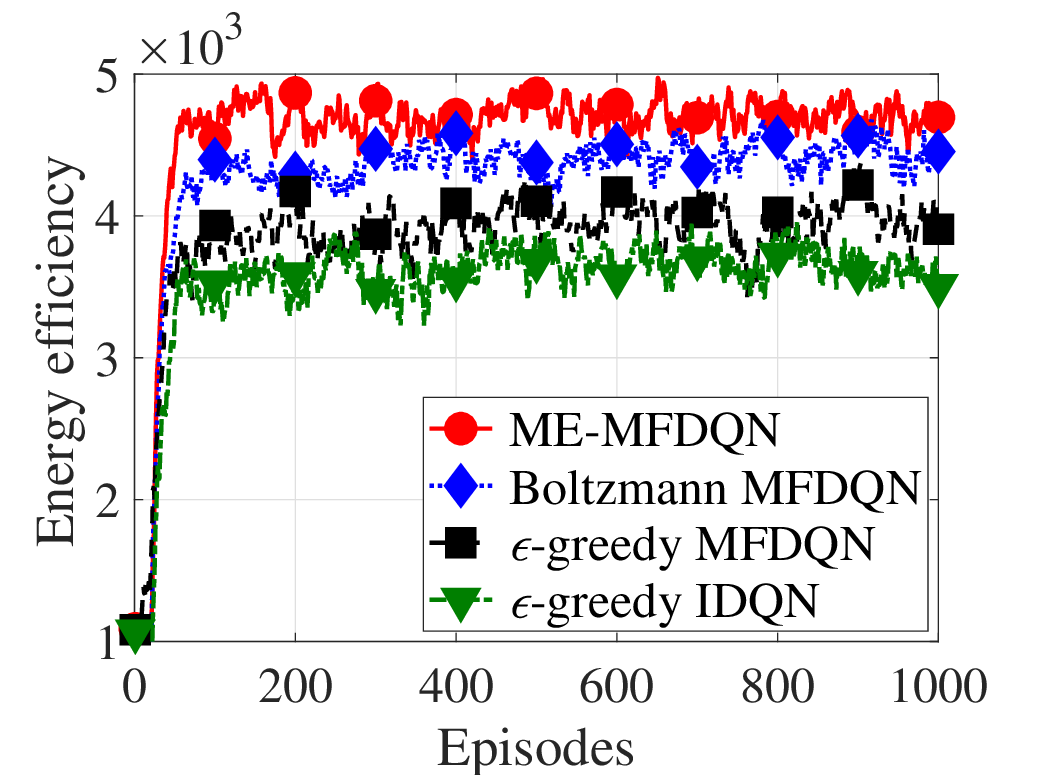}}\hspace{-3mm}
	\subfigure[]{
		\includegraphics[width=2.2in]{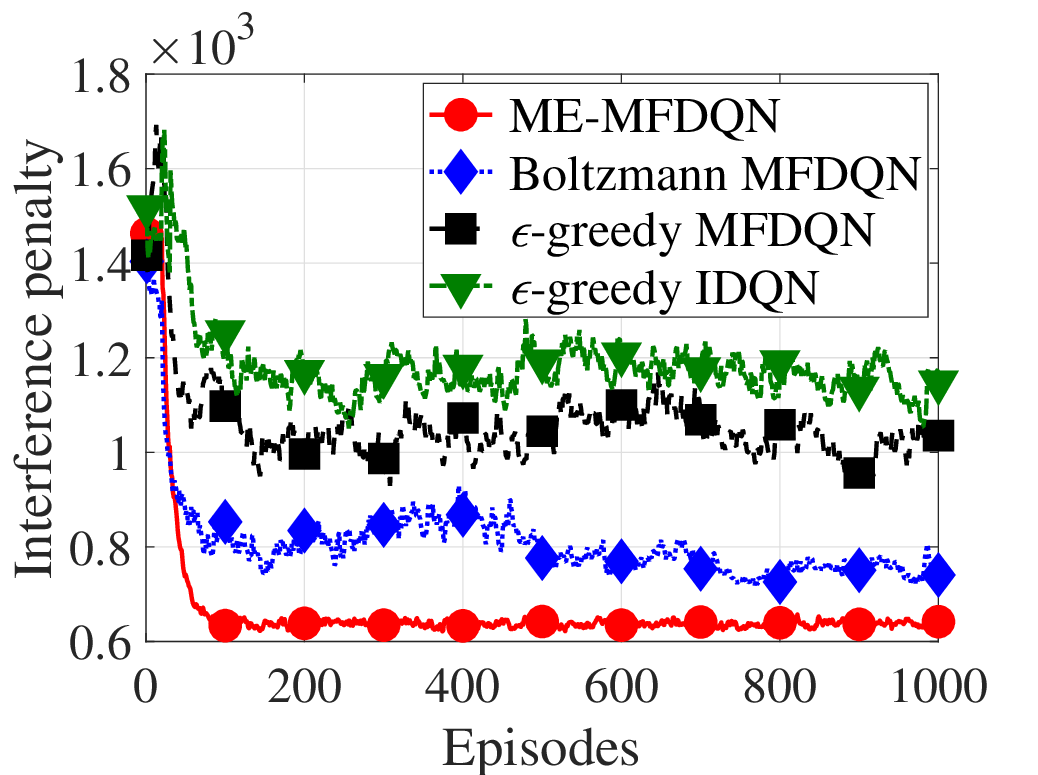}}
    \caption{Average reward, energy efficiency and interference penalty per episode received by the representative agent under different MFRL algorithms: (a) average reward; (b) energy efficiency; (c) interference penalty.}
    \hrule
    \vspace{-3mm}
  \end{figure*}

  In this section, we present the simulation results to evaluate the policy and performance (energy efficiency) of UAVs under the proposed ME-MFDQN algorithm for the fully and partially observable scenarios.
  The communication bandwidth is $B = 1$ $\rm{MHz}$.
  The UAV altitude is set as~$d = 100$~$\rm{m}$. UAV $k$'s transmit power is $p_k(t) \in$ \{$0$~$\rm{mW}$, $50$~$\rm{mW}$, $100$~$\rm{mW}$, $150$~$\rm{mW}$, $200$~$\rm{mW}$\} and the on-board circuit power~$p^{\rm{c}}$ is~$0.01$~$\rm{W}$. The noise power $N_0$ is~$-110$~$\rm{dBm}$~\cite{AD_UAV}.
  According to~\cite{Optimal_3D}, the parameters of the solar energy harvesting model are set as $\varrho = 0.4$, $\vartheta = 1$~${\rm{m}}^2$, $G = 1367$~${\rm{W/m}}^2$, $\beta = 0.01$, and $d^{\rm{cloud}}(t) \in$ \{$0$ $\rm{m}$, $700$ $\rm{m}$\}.
  The modeling parameters for the LoS probability are $c_1 = 10$ and~$c_2 = 0.6$.
  According to~\cite{Down_cov} and \cite{Pro_caching}, the parameters of the channel model are listed in Table~I. According to \cite{Energy_Min}, the corresponding parameters for the UAV's propulsion power consumption model are specified in Table~II.
  We set the number of UAVs to $19\times19$ with the density of $1/{\rm{km}}^2$, where each UAV serves $4$ GUs. The representative UAV is located at the center of the network. The lengths of~$\tau$, $\tau_1$ and~$\tau_2$ are set as $60$ $\rm{s}$, $25$~$\rm{s}$ and~$35$~$\rm{s}$, respectively.
  Under the above parameter settings, the flying energy consumption of the UAV is always greater than the hovering energy consumption.
  Unless otherwise stated, the SINR threshold~$\eta$ is set as $0$ $\rm{dB}$, the self-transition probability $q$ is $0.7$, and the penalty factor $\sigma$ is $240$.

  \begin{table}[H]
  \newcommand{\tabincell}[2]{\begin{tabular}{@{}#1@{}}#2\end{tabular}}
  \renewcommand{\arraystretch}{1.4}
  \centering
  \caption{List of channel model parameters}
  \begin{tabular}{|c|c|c|c|}
    \hline
    Parameter       &Value    &Parameter       &Value                               \\  \hline
    $A^{\rm{L}}$    &$10^{-3.692}$   &$\alpha^{\rm{L}}$    &$2.225-0.05\log_{10}(d)$         \\  \hline
    $A^{\rm{NL}}$   &$10^{-3.842}$   &$\alpha^{\rm{NL}}$   &$4.32-0.76\log_{10}(d)$            \\  \hline
    $m$                  &$2$       &$\Omega$          &$1$                            \\  \hline
  \end{tabular}
  \end{table}

  \vspace{-2mm}
  \begin{table}[H]
  \newcommand{\tabincell}[2]{\begin{tabular}{@{}#1@{}}#2\end{tabular}}
  \renewcommand{\arraystretch}{1.4}
  \centering
  \caption{List of UAV's propulsion power parameters}
  \begin{tabular}{|c|c|c|c|}
    \hline
    Parameter       &Value    &Parameter       &Value                               \\  \hline
    $W$   &$20$ $\rm{Newton}$   &$\rho$   & $1.225$ ${\rm{kg/m}}^3$     \\  \hline
    $\omega^{\rm{rad}}$   & $0.4$ $\rm{m}$   &$\omega^{\rm{area}}$   & $0.503$ ${\rm{m}}^2$             \\  \hline
    $\omega^{\rm{sol}}$   &$0.05$  &$\zeta$  & $300$ ${\rm{radians/s}}$                             \\  \hline
    $\varepsilon$   &$0.6$ &$\varpi$   &$0.012$                          \\  \hline
    $\lambda$  &$0.1$ & &               \\  \hline
  \end{tabular}
  \end{table}

  \subsection{Experimental Evaluation for ME-MFDQN}
  We implement the proposed ME-MFDQN algorithm in the large-scale multi-UAV network for the fully observable scenario, where each UAV is able to observe the real-time service demands for all GUs within its cell.
  We set the total number of training iterations as $1000$, where there are $200$ steps for each iteration.
  We set the discount factor as $\gamma = 0.9$.
  The neural network is trained by the Adam optimizer, and the learning rate of the Q network is set to~$0.005$.
  There are two hidden layers in this Q network, where there are $128$ and $64$ neurons for the first and second layers, respectively.
  At each time slot, the representative agent can randomly sample a mini-batch with $300$ experiences from the replay buffer which can store up to $1000$ past experiences.

  In Fig.~4, we evaluate the performance of the proposed ME-MFDQN algorithm in terms of the average reward in Fig. 4(a), and further decompose it into energy efficiency in Fig. 4(b) and interference penalty in Fig. 4(c). We further propose three benchmark algorithms, i.e., $\epsilon$-greedy independent DQN (IDQN), $\epsilon$-greedy MFDQN and Boltzmann MFDQN, respectively.
  For $\epsilon$-greedy IDQN, each UAV independently interacts with the environment by using a single-agent DQN algorithm with the $\epsilon$-greedy policy.
  Since the UAV does not directly model the behavior of other agents, the performance is the worst, i.e., with lowest reward and energy efficiency as shown in Figs. 4(a) and 4(b), and highest interference penalty in Fig. 4(c).
  To further characterize the interactions among UAVs' policies, we extend $\epsilon$-greedy IDQN to $\epsilon$-greedy MFDQN, where each UAV not only adopts DQN algorithm to obtain its best response, but also updates the collective behavior of all other UAVs via mean-field distributions. Though $\epsilon$-greedy MFDQN outperforms the independent counterpart, both methods adopt $\epsilon$-greedy policy, where the exploration efficiency is limited due to the fixed exploration probability~$\epsilon$.
  For this reason, we introduce Boltzmann MFDQN to enable the UAV to update the policies according to the Boltzmann probability, i.e., using higher probability to select the actions with higher Q values.
  Due to the~better exploration capability, the performance of Boltzmann MFDQN outperforms $\epsilon$-greedy MFDQN.
  In order to further improve the exploration efficiency, our proposed ME-MFDQN increases the randomness of the policy selection by maximizing the entropy of the policy as a part of the reward function.
  The results demonstrate that ME-MFDQN outperforms the above three benchmark algorithms. Due to the higher exploration efficiency of this algorithm, UAVs can select the actions more effectively, resulting in higher energy efficiency and lower interference penalty as shown in Figs.~4(b) and~4(c).

  \begin{figure}[t]
    \centering
    \includegraphics[width=2.5in]{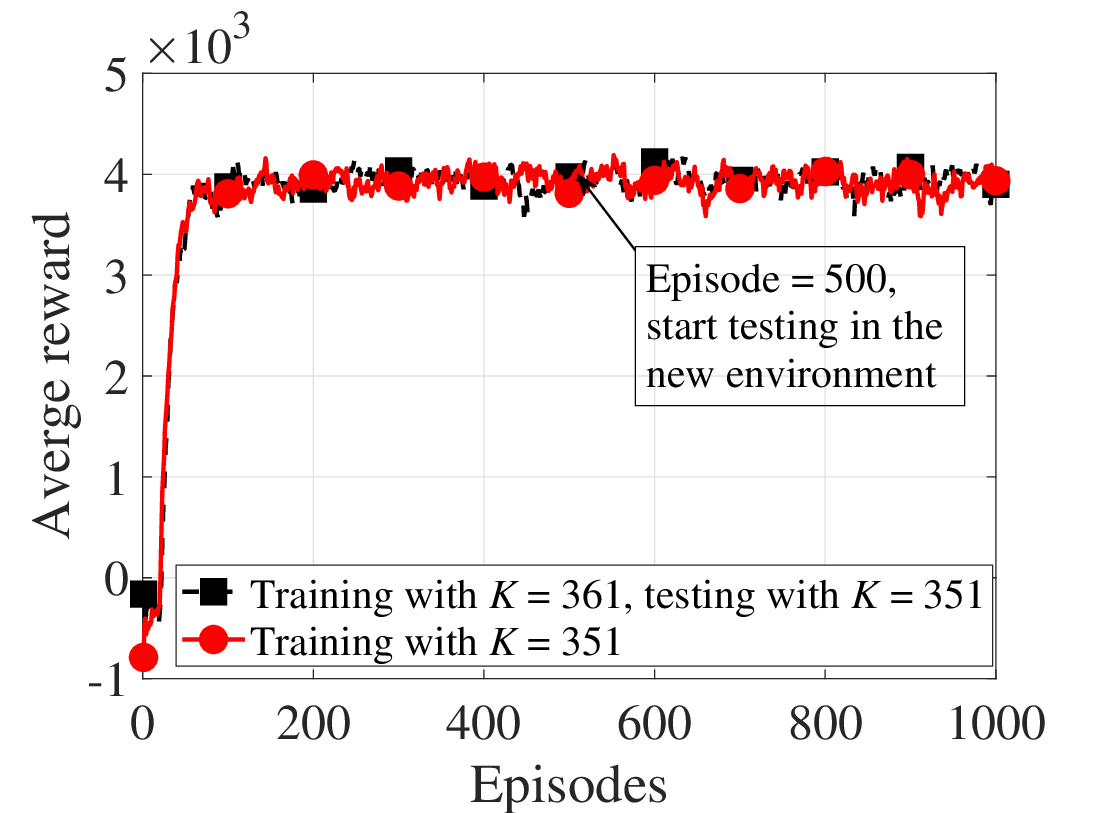}
    \caption{Average reward per episode for the active UAVs.}
    \vspace{-2mm}
  \end{figure}

  In Fig. 5, we validate the robustness of the proposed algorithm under different UAV numbers. As for most literature, the RL policy should be retrained from the scratch once the UAV number changes.
  To address this issue, our recent work in~\cite{DT} proposed a decision transformer based approach to enable the policy generalization across different numbers of UAVs or users. However, the slight variation of UAV numbers may not much affect the large-scale system, where the impact of an individual agent on the aggregate behavior of the mean-field is limited. To verify this, we first train the UAV policies via the proposed algorithm in a scenario with $361$ UAVs, where it converges after $200$ episodes. Then, we directly apply the trained policy in a new environment with $351$ UAVs right after the $500$-th episode, assuming that $10$ UAVs are damaged or disconnected from the network.
  It shows that the average reward of the active UAVs keeps almost unchanged despite the variations on the UAV numbers, indicating the robustness of the proposed algorithm.

  \begin{figure}[t]
	\centering
	\subfigbottomskip=2pt
	\subfigure[]{
		\includegraphics[width=1.66in]{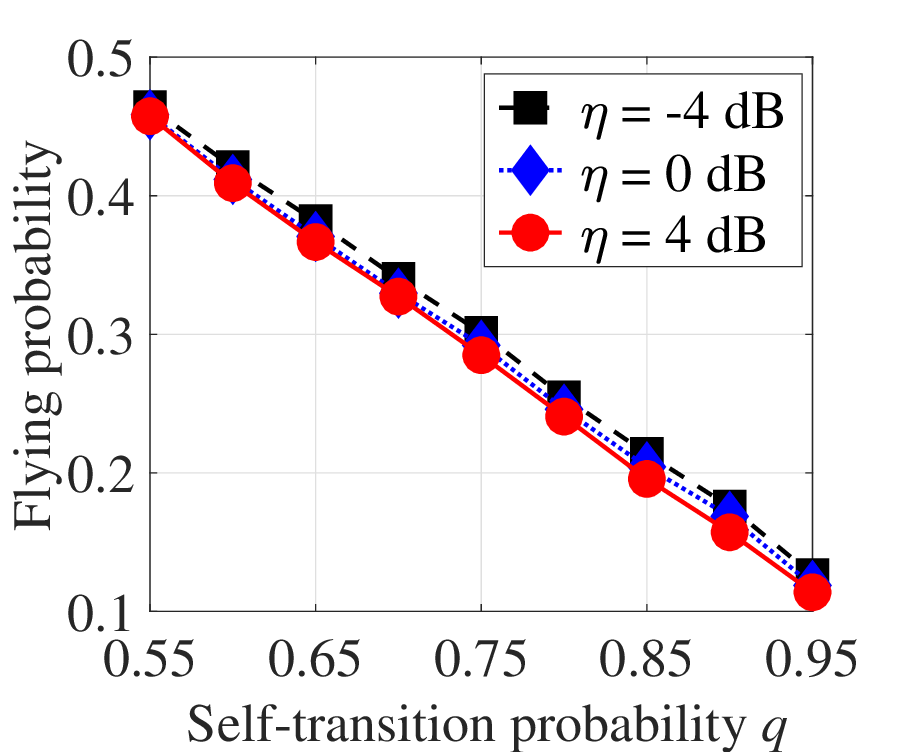}}\hspace{-3mm}
	\subfigure[]{
		\includegraphics[width=1.66in]{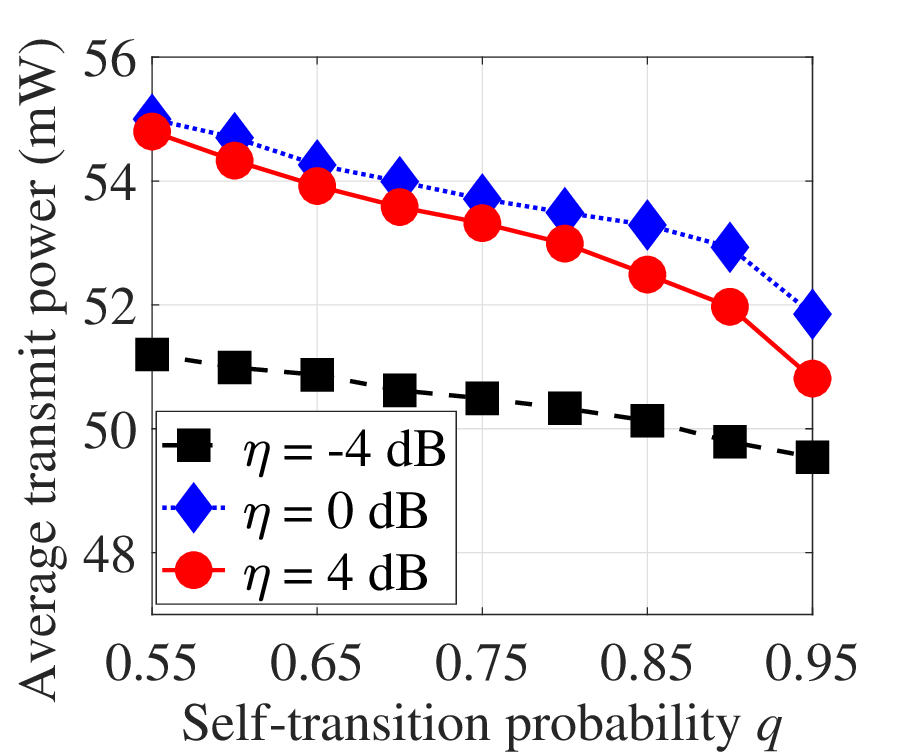}}
    \caption{The impact of self-transition probability $q$ on the optimal policies: (a) the flying probability versus $q$; (b) the average transmit power versus $q$.}
    \vspace{-2mm}
  \end{figure}

  \begin{figure}[t]
    \centering
    \includegraphics[width=2.5in]{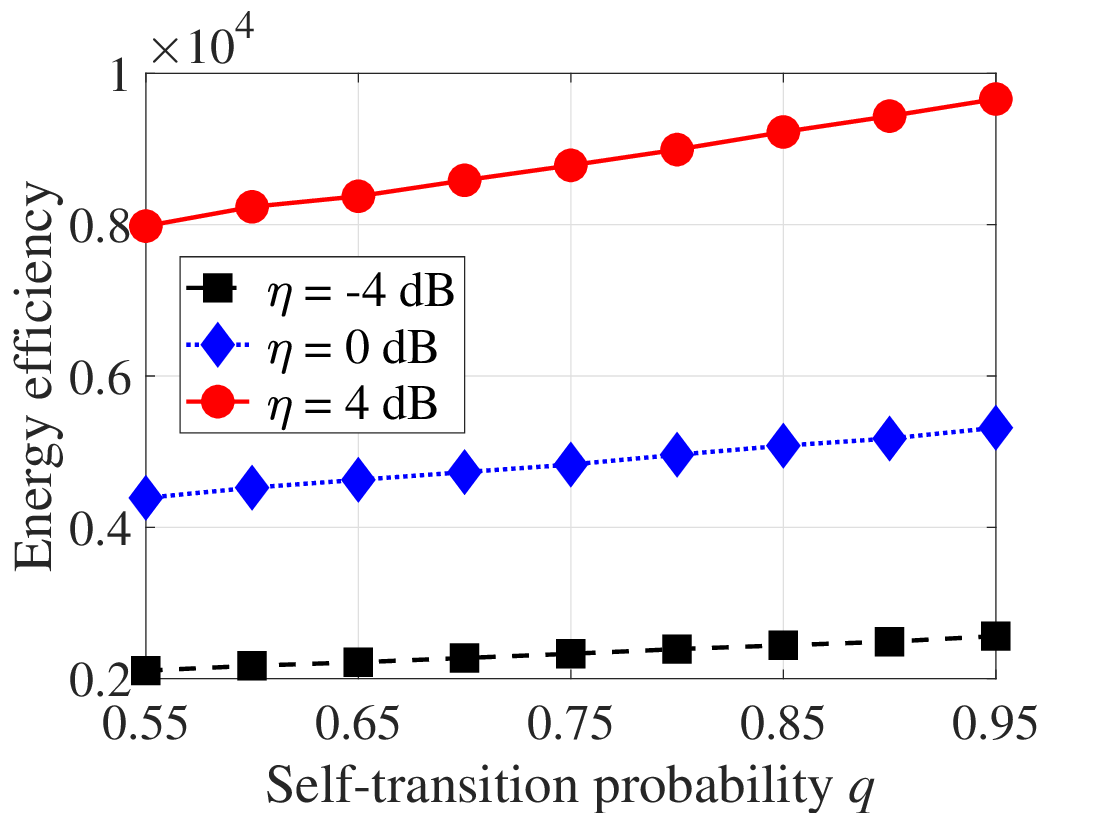}
    \caption{Energy efficiency versus various self-transition probability $q$ under different SINR thresholds $\eta$.}
    \vspace{-2mm}
  \end{figure}

  Fig.~6 plots the optimal policies of the representative UAV against the self-transition probability $q$ of the GUs' service demands. We use the flying probability to represent the probability that the UAV chooses a new hovering point at the beginning of the time slot. It shows that both the flying probability and the average transmit power of the representative UAV decrease with $q$. When $q$ is small, the GUs' demands are highly random, and the active GU served by the UAV is likely to turn idle in the next time step.
  In this case, the UAV is more likely to fly away to find new service opportunities (as shown in Fig.~6(a)), and may also choose a higher power for data transmission to enhance the service coverage (as shown in Fig.~6(b)).
  When $q$ is large, the GUs' demands change slowly.
  In this case, the UAV flies less frequently to find new active GUs, and requires lower power to serve local GUs.
  In addition, with the increase of SINR threshold $\eta$, the flying probability does not change much~(as shown in Fig.~6(a)), and the average transmit power first increases and then decreases~(as shown in Fig.~6(b)).
  We discuss the influence of SINR threshold $\eta$ on transmit power as follows.
  According to~\eqref{reward}, the reward function can be divided into two terms: energy efficiency and interference penalty.
  Take~$q=0.95$ as an example.
  When the SINR threshold $\eta$ is small ($-4$~$\rm{dB}$), the impact of the achievable data rate is smaller than that of the interference penalty, and thus the UAV adopts low power~($49.5$~$\rm{mW}$) to avoid excessive interference penalty.
  When $\eta$ increases to~$0$~$\rm{dB}$, the impact of the energy efficiency reward increases, and thus the UAV adopts a high power ($52$ $\rm{mW}$) to enhance the coverage.
  When $\eta$ increases to $4$ $\rm{dB}$, a higher power is desirable for energy efficiency enhancement, but it is very costly due to the large interference penalty.
  In this case, the UAV would rather choose a relatively low power ($51$ $\rm{mW}$) for interference avoidance.

  Fig.~7 shows that the energy efficiency of the representative UAV increases with the increase of the self-transition probability $q$ of the GUs' service demands.
  When $q$ is small, the UAV is uncertain about the GUs' demands and will choose a relatively high flying probability and high transmit power (as shown in Fig.~6), which increases its time/energy cost and thus reduces the energy efficiency.
  When $q$ is large, the UAV is more certain about the GUs' demands and will adopt a smaller flying probability and transmit power~(as shown in Fig.~6). In this case, the UAV has a lower time/energy cost, and thus improves its energy efficiency.
  In addition, it can be seen from Fig.~7 that the UAV's energy efficiency increases with SINR threshold $\eta$ due to the higher achievable data rate.

  Figs. 8(a) and 8(b) show that with the increase of power penalty factor $\sigma$, the representative UAV's flying probability increases and its average transmit power decreases, respectively.
  When $\sigma$ is small, the interference penalty term in \eqref{reward} has limited impact, and thus the UAV can choose a relatively high transmit power to improve its throughput (as shown in Fig.~8(b)), while choosing a low flying probability for cost reduction (as shown in Fig.~8(a)).
  As $\sigma$ gradually increases, the influence of the interference penalty item increases. Thus, the UAV needs to fly closer to the potentially active GUs (as shown in Fig.~8(a)) while transmitting at a lower power (as shown in Fig.~8(b)) for interference control purposes.
  In addition, with the increase of the SINR threshold~$\eta$, the flying probability increases monotonously, but its impact on the average transmit power is more complex.
  Firstly, when the penalty factor is small~($\sigma \leq 320$), the average transmit power increases first and then decreases with the increase of $\eta$, which is consistent with the results shown in Fig.~6(b).
  Secondly, with the increase of penalty factor ($320 < \sigma \leq 800$), the average transmit power increases first and then remains almost unchanged with the increase of $\eta$.
  Moreover, we find that it is not always wise to use a large penalty factor since this is costly for the UAV to frequently fly closer to serve GUs at low power for interference control.
  Therefore, designing an appropriate power penalty factor is crucial for improving the system performance.

%

    \begin{figure}[t]
	\centering
	\subfigbottomskip=2pt
	\subfigure[]{
		\includegraphics[width=1.66in]{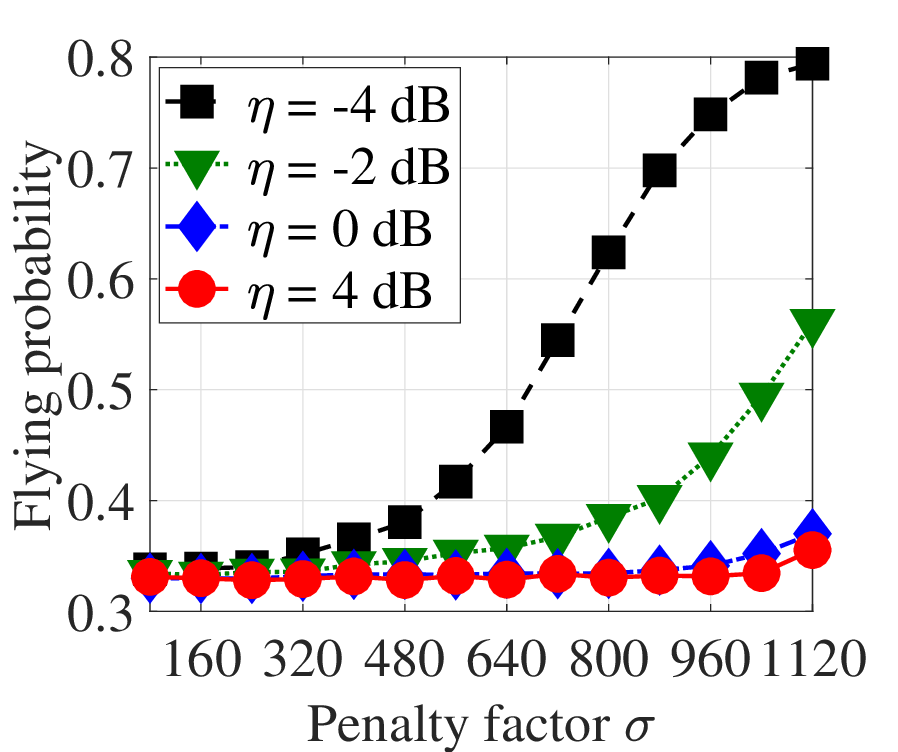}}\hspace{-3mm}
	\subfigure[]{
		\includegraphics[width=1.66in]{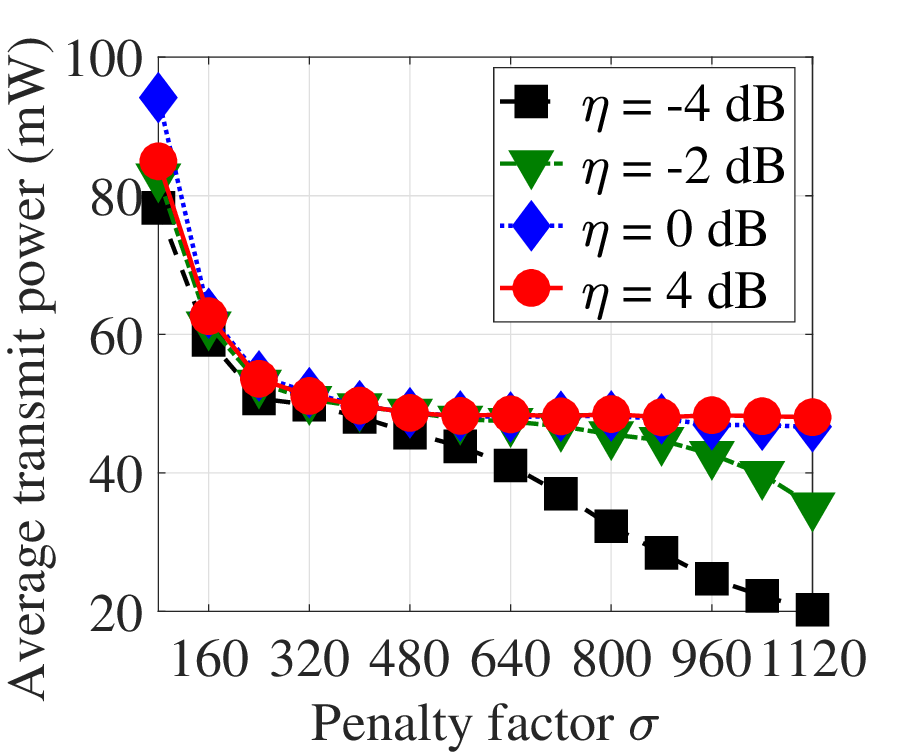}}
    \caption{The impact of penalty factor $\sigma$ on the optimal policies: (a) the flying probability versus $\sigma$; (b) the average transmit power versus $\sigma$.}
    \vspace{-2mm}
  \end{figure}

  \begin{figure}[t]
    \centering
    \includegraphics[width=3.5in,angle=0]{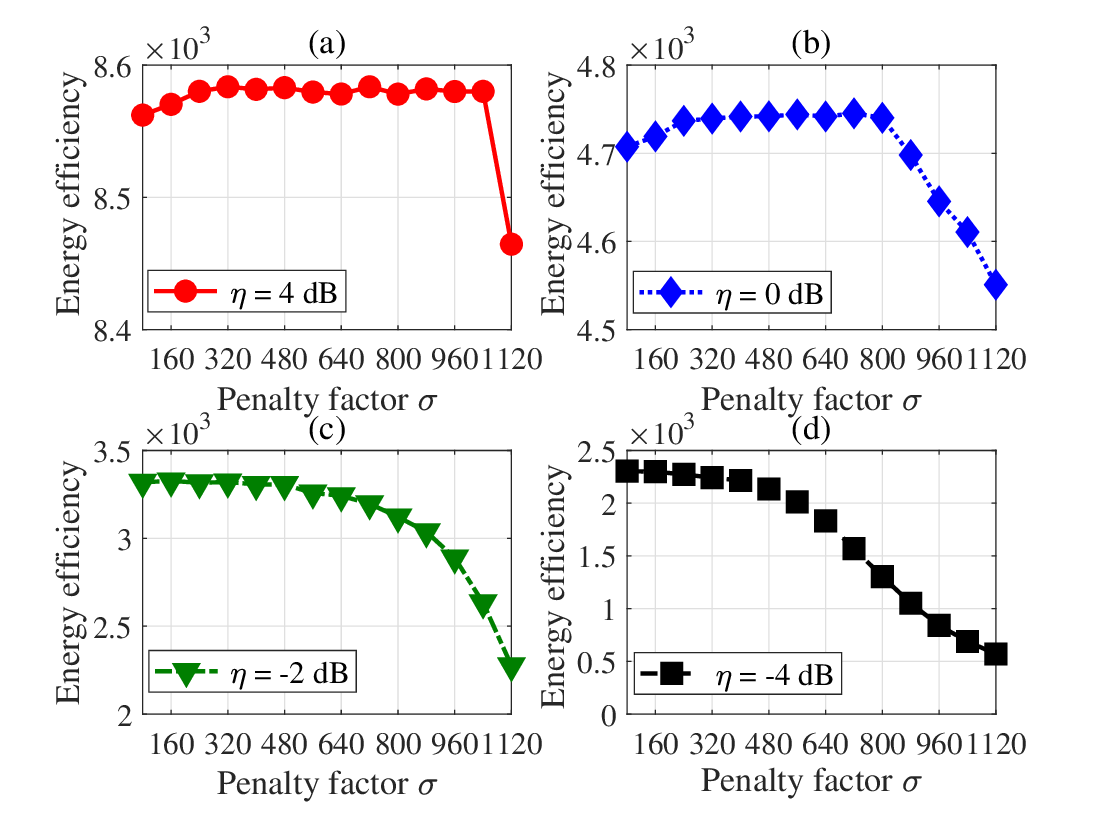}
    \caption{Energy efficiency versus various penalty factor $\sigma$ under different SINR thresholds $\eta$.}
    \vspace{-2mm}
  \end{figure}

  Fig.~9 shows the impact of the penalty factor $\sigma$ on the energy efficiency of the representative UAV.
  When the SINR thresholds $\eta$ is relatively large ($0$ $\rm{dB}$ and $4$ $\rm{dB}$) as shown in Figs.~9(a) and 9(b), the energy efficiency first increases, then remains static, and finally decreases with $\eta$.
  Firstly, when the interference penalty factor $\sigma$ is small, the UAV will choose a large transmit power (as shown in Fig.~8(b)), resulting in high interference, and thus has low energy efficiency.
  Secondly, for a high $\sigma$ regime, e.g., $[320,1040]$ for $4$ $\rm{dB}$ and $[240,800]$ for $0$ $\rm{dB}$,
  the UAV' energy efficiency remains almost unchanged since its policy does not change much within these regions (as shown in Fig.~8).
  Finally, with the further increase of $\sigma$, the UAV will increase the flying probability while reducing the transmit power for interference control (as shown in Fig.~8), which will increase the time/energy cost and thus reduce the energy efficiency.
  When the SINR threshold $\eta$ is relatively small~($-2$ $\rm{dB}$ and $-4$~$\rm{dB}$) as shown in Figs.~9(c) and 9(d), the energy efficiency curves follow a similar trend as the above two subfigures except for the low $\sigma$ regime, since it is more tolerable to the interference when the target rate is low.
  In addition, we can see that the desirable penalty factor regions are $[320, 1040]$ for $\eta = 4$ $\rm{dB}$, $[240, 880]$ for~$\eta = 0$~$\rm{dB}$, $[80, 480]$ for $\eta = -2$ $\rm{dB}$, and $[80, 320]$ for $\eta = -4$ $\rm{dB}$, respectively.

  \subsection{Experimental Evaluation for Partially observable ME-MFDQN}

  Fig.~10 compares the average reward of the proposed ME-MFDQN algorithm between the fully and partially observable scenarios.
  We consider the following three typical cases: each UAV can fully ($100\%$) observe the service demands of all GUs in its cell, or partially observe that of the nearest~$75\%$ or $25\%$ GUs in the cell.
  The proposed algorithm shows good convergence in all three cases.
  The results show that the average reward of the representative agent increases with more observation information.
  The average reward of the fully observable case slightly outperforms that of $75\%$ observation, where both are superior to the case of~$25\%$ observation.

  \begin{figure}[t]
    \centering
    \includegraphics[width=2.5in,angle=0]{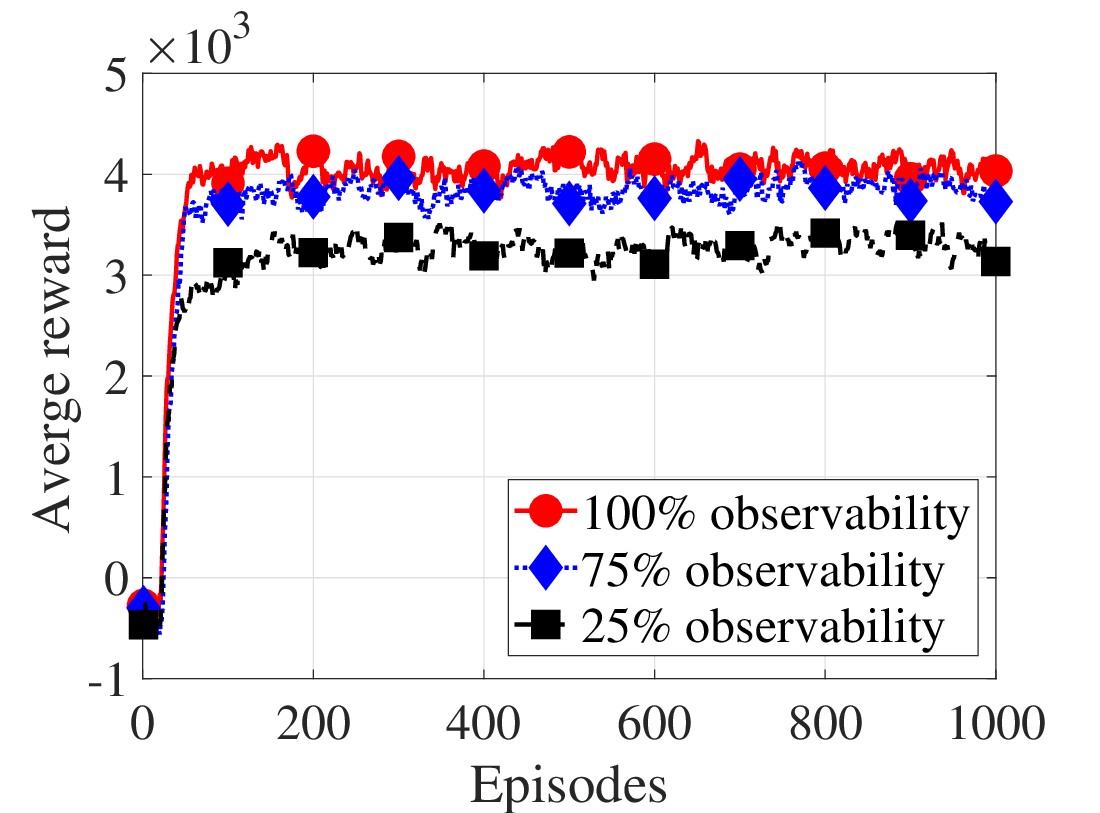}
    \caption{Average reward per episode received by the representative agent under fully and partially observable scenarios.}
    \vspace{-2mm}
  \end{figure}

  Fig.~11 further compares the effect of $q$ on the flying probability and average transmit power in both the fully and partially observable scenarios.
  We observe that both the flying probability and average transmit power decreases with the increase of $q$, where the reasons are similar to that of Fig.~6 and are omitted here. Moreover, Fig.~11(a) shows that the flying probability decreases with more observations on the GUs' demands.  With a larger observation radius, the UAV is more certain about the environment and is thus less necessary to frequently flying around to explore the new opportunities. Furthermore, Fig.~11(b) demonstrates that the average transmit power first increases and then decreases with the increase of observation radius. Particularly, we observe that the average transmit power of the UAV with $75\%$ observability is even greater than that of the full observation case. This is because the UAV intends to compensate for the lack of full observation at the cost of a higher transmit power, in order to improve the local energy efficiency. For the UAV with $25\%$ observability, it is safer to keep a relatively low transmit power to avoid excessive interference penalty when it is less certain about the environment.


  \begin{figure}[t]
	\centering
	\subfigbottomskip=2pt
	\subfigure[]{
		\includegraphics[width=1.66in]{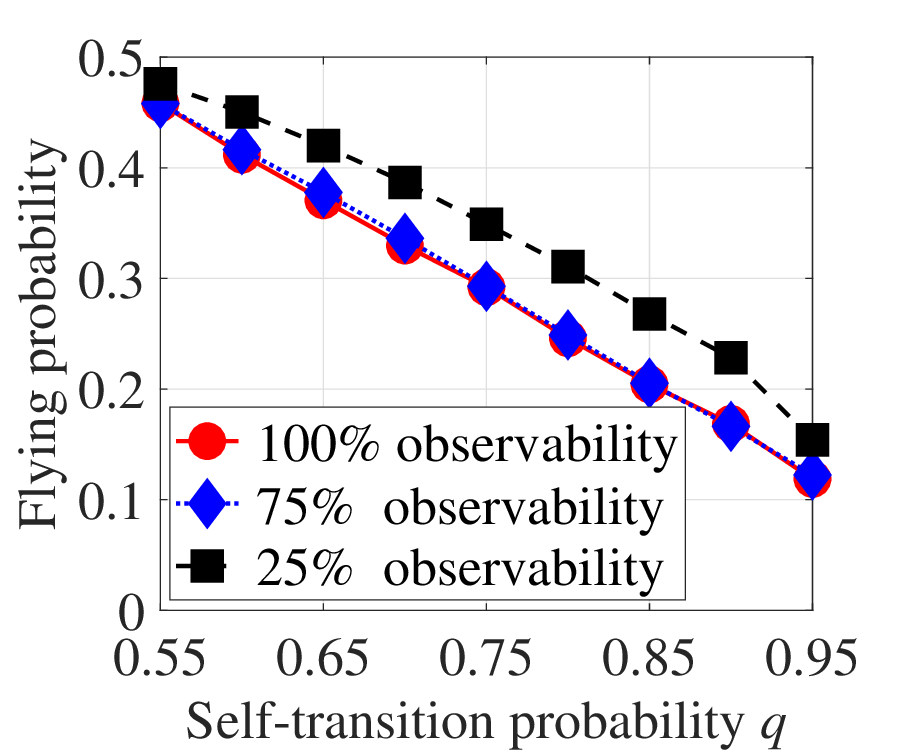}}\hspace{-3mm}
	\subfigure[]{
		\includegraphics[width=1.66in]{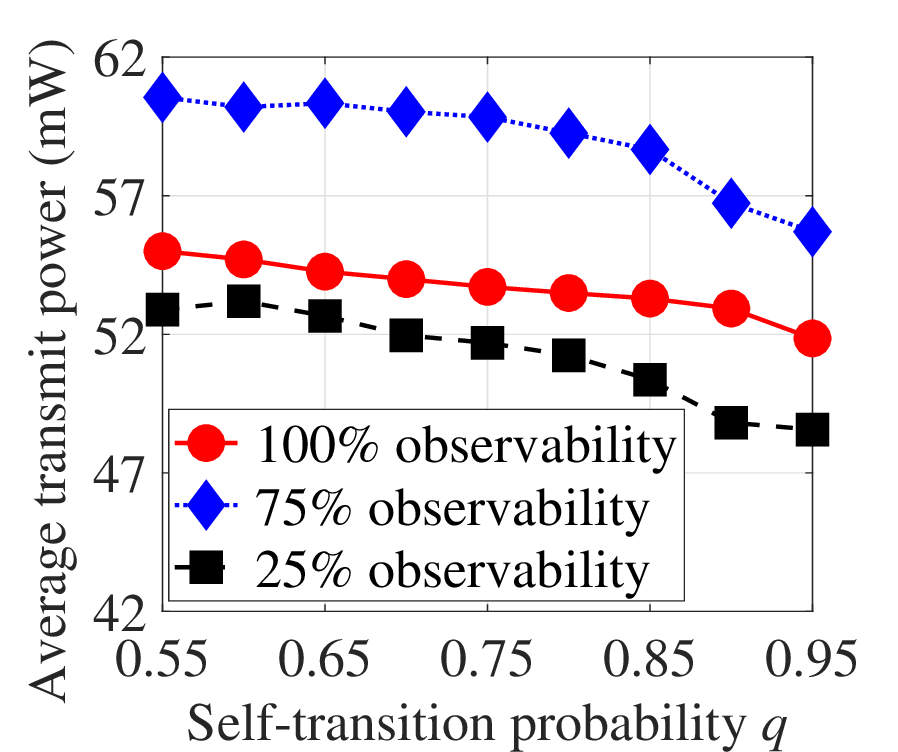}}
    \caption{The impact of self-transition probability $q$ under fully and partially observation scenarios: (a) the flying probability versus $q$; (b) the average transmit power versus~$q$.}
    \vspace{-2mm}
  \end{figure}

  \begin{figure}[t]
    \centering
    \includegraphics[width=2.5in,angle=0]{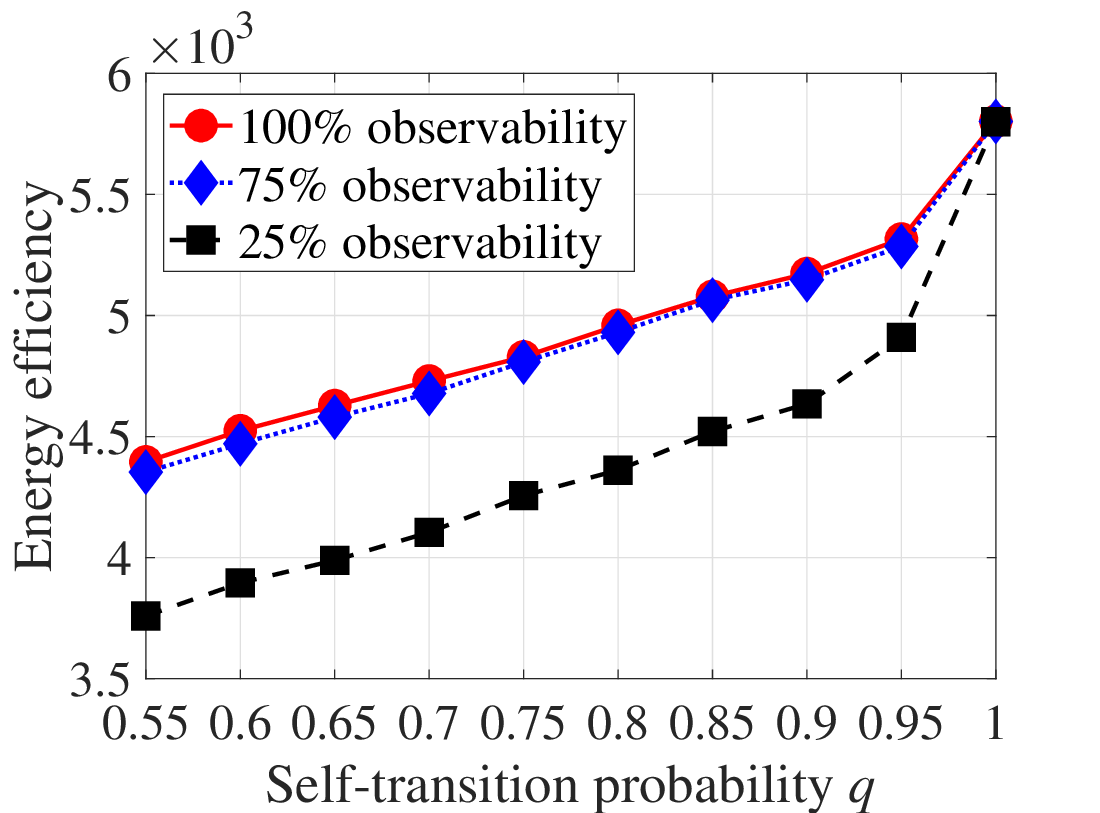}
    \caption{Energy efficiency versus various self-transition probability $q$ in fully and partially observation scenarios.}
    \vspace{-2mm}
  \end{figure}

  Fig.~12 illustrates that the average energy efficiency for the representative UAV increases with $q$ for both the fully and partially observable scenarios.
  We can see that the UAV achieves similar energy efficiency in $75\%$ and $100\%$ observability cases since it has similar flying policies in these two scenarios (as shown in Fig.~11(a)).
  Though the transmit power is significantly higher in the $75\%$ observability case than that of full observation case (as shown in Fig.~11(b)), it does not have much influence on the energy efficiency since the transmit power is almost ignorable compared with the flying power in the total energy consumption. Furthermore, in $25\%$ observability case, it has significantly less energy efficiency than the other two cases since it needs to frequently flying across the users to compensate for the extreme lack of observation information.

  \section{Conclusion}

  In this paper, we proposed a non-cooperative resource allocation scheme in an ultra-dense UAV communication network for accommodating dynamic GUs' service demands. We formulated the interactions among the co-channel UAVs as an MFG, where each self-interested UAV aims to maximize the expectation of its locally cumulative energy efficiency under interference and energy constraints by jointly optimizing its transmit power, user association and trajectory. We proved the existence of mean-field equilibrium and proposed the ME-MFDQN algorithm to solve the mean-field equilibrium in both fully and partially observable scenarios. The simulation results demonstrated the effectiveness of the proposed ME-MFDQN algorithm in enhancing the energy efficiency compared with the benchmark algorithms.
  Furthermore, we observed that if the GUs' service demands exhibit higher temporal correlation, the UAVs would decrease their transmit power and flying probability, for improving the energy efficiency.
  In addition, energy efficiency was further enhanced given a wider observation radius of the UAV.

  \begin{appendices}

  \section{ Proof of Theorem 1 }

  \begin{definition}
  \label{de_appe}
  For any policies $\bm{\pi}, \hat{\bm{\pi}} \in \Pi$ and any distributions $\bm{\mathcal{L}}, \hat{\bm{\mathcal{L}}} \in \mathcal{M}$, we define the metric spaces $(\Pi, d_{\Pi})$ and~$(\mathcal{M}, d_{\mathcal{M}})$ as
  \begin{align}
  &d_{\Pi}(\bm{\pi}, \hat{\bm{\pi}}):= \sup_{t\in \mathcal{T}}\sup_{s(t)\in \mathcal{S}} d_{\rm{W}}(\pi(\cdot|s(t)), \hat{\pi}(\cdot|s(t))),
  \end{align}
  and
  \begin{align}
  &d_{\mathcal{M}}(\bm{\mathcal{L}}, \hat{\bm{\mathcal{L}}}):= \sup_{t\in \mathcal{T}} d_{\rm{W}}(\mathcal{L}(t), \hat{\mathcal{L}}(t)),
  \end{align}
  where $d_{\rm{W}}$ is the $\ell_1$-Wasserstein distance between probability measures \cite{LMFG}.
  \end{definition}

  \begin{lemma}
  For any distributions $\bm{\mathcal{L}}, \hat{\bm{\mathcal{L}}} \in \mathcal{M}$, there exists a constant $\varsigma_1 \geq 0$, such that
  \begin{align}
  \label{assup1}
  d_{\Pi}(\Upsilon_1(\bm{\mathcal{L}}), \Upsilon_1(\hat{\bm{\mathcal{L}}})) \leq \varsigma_1 d_{\mathcal{M}}(\bm{\mathcal{L}}, \hat{\bm{\mathcal{L}}}).
  \end{align}
  \end{lemma}
  The proof of Lemma 1 is given in \cite{LP_SG} and is omitted here.

  \begin{lemma}
  For any policies $\bm{\pi}, \hat{\bm{\pi}} \in \Pi$ and any distributions $\bm{\mathcal{L}}, \hat{\bm{\mathcal{L}}} \in \mathcal{M}$, there exists constants $\varsigma_2, \varsigma_3 \geq 0$, such that
  \begin{align}
  \label{assup2_1}
  d_{\mathcal{M}}(\Upsilon_2(\bm{\pi}, \bm{\mathcal{L}}), \Upsilon_2(\hat{\bm{\pi}}, \bm{\mathcal{L}})) \leq \varsigma_2 d_{\Pi}(\bm{\pi}, \hat{\bm{\pi}}),
  \end{align}
  and
  \begin{align}
  \label{assup2_2}
  d_{\mathcal{M}}(\Upsilon_2(\bm{\pi}, \bm{\mathcal{L}}), \Upsilon_2(\bm{\pi}, \hat{\bm{\mathcal{L}}})) \leq \varsigma_3 d_{\mathcal{M}}(\bm{\mathcal{L}}, \hat{\bm{\mathcal{L}}}).
  \end{align}
  \end{lemma}
  The proof of Lemma 2 is given in \cite{LMFG} and is omitted here.

  Based on the above definition and lemmas, $(\bm{\pi}^*_{\bm{\mathcal{L}}}, \bm{\mathcal{L}})$ is a stationary mean-field equilibrium if and only if $\bm{\pi}^*_{\bm{\mathcal{L}}} = \Upsilon_1(\bm{\mathcal{L}})$, $\Upsilon(\bm{\mathcal{L}}) = \Upsilon_2(\Upsilon_1(\bm{\mathcal{L}}), \bm{\mathcal{L}})$.
  For any $\bm{\mathcal{L}}, \hat{\bm{\mathcal{L}}} \in \mathcal{ M}$, there exists
  \begin{align}
  d_{\mathcal{M}}(&\Upsilon(\bm{\mathcal{L}}), \Upsilon(\hat{\bm{\mathcal{L}}})) \nonumber \\
  & = d_{\mathcal{M}}(\Upsilon_2(\Upsilon_1(\bm{\mathcal{L}}), \bm{\mathcal{L}}), \Upsilon_2(\Upsilon_1(\hat{\bm{\mathcal{L}}}), \hat{\bm{\mathcal{L}}})) \nonumber\\
  & \mathop \leq \limits^{(a)} d_{\mathcal{M}}(\Upsilon_2(\Upsilon_1(\bm{\mathcal{L}}), \bm{\mathcal{L}}), \Upsilon_2(\Upsilon_1(\hat{\bm{\mathcal{L}}}), \bm{\mathcal{L}})) \nonumber\\
  &\;\;\;\;+ d_{\mathcal{M}}(\Upsilon_2(\Upsilon_1(\hat{\bm{\mathcal{L}}}), \bm{\mathcal{L}}), \Upsilon_2(\Upsilon_1(\hat{\bm{\mathcal{L}}}), \hat{\bm{\mathcal{L}}})) \nonumber\\
  & \mathop \leq \limits^{(b)} \varsigma_2 d_{\Pi}(\Upsilon_1(\bm{\mathcal{L}}), \Upsilon_1(\hat{\bm{\mathcal{L}}})) + \varsigma_3 d_{\mathcal{M}}(\bm{\mathcal{L}}, \hat{\bm{\mathcal{L}}}) \nonumber \\
  & \mathop \leq \limits^{(c)} \varsigma_2 \varsigma_1 d_{\mathcal{M}}(\bm{\mathcal{L}}, \hat{\bm{\mathcal{L}}}) + \varsigma_3 d_{\mathcal{M}}(\bm{\mathcal{L}}, \hat{\bm{\mathcal{L}})} \nonumber \\
  & = (\varsigma_1 \varsigma_2+\varsigma_3) d_{\mathcal{M}}(\bm{\mathcal{L}}, \hat{\bm{\mathcal{L}}}),
  \end{align}
  where step (a) holds because of the $\ell_1$-Wasserstein distance between $\Upsilon_2(\Upsilon_1(\bm{\mathcal{L}}), \bm{\mathcal{L}})$ and~$\Upsilon_2(\Upsilon_1(\bm{\mathcal{L}}), \hat{\bm{\mathcal{L}}})$ plus the $\ell_1$-Wasserstein distance between $\Upsilon_2(\Upsilon_1(\bm{\mathcal{L}}), \hat{\bm{\mathcal{L}}})$ and~$\Upsilon_2(\Upsilon_1(\hat{\bm{\mathcal{L}}}), \hat{\bm{\mathcal{L}}})$ is greater than or equal to the~$\ell_1$-Wasserstein distance between $\Upsilon_2(\Upsilon_1(\bm{\mathcal{L}}), \bm{\mathcal{L}})$ and~$\Upsilon_2(\Upsilon_1(\hat{\bm{\mathcal{L}}}), \hat{\bm{\mathcal{L}}})$. Step (b) holds due to \eqref{assup2_1} and \eqref{assup2_2}, and step (c) holds because of \eqref{assup1}.

  Based on the Banach fixed point theorem given in \cite{Appr_sol} and \cite{LMFG}, it can converge to a unique fixed point of $\Upsilon$, that is, a unique mean-field equilibrium solution.

  \end{appendices}

\bibliographystyle{IEEEtran}
\bibliography{reference}

\end{document}